# Development of structural descriptors to predict dissolution rate of volcanic glasses: molecular dynamic simulations

[1]Kai Gong and [1*]Elsa Olivetti

[1]Department of Materials Science and Engineering, Massachusetts Institute of Technology, Cambridge, MA 02139, USA

*Correspondence: Elsa A. Olivetti, Department of Materials Science and Engineering, MIT, Cambridge 02139, USA. Email: elsao@mit.edu

## Abstract

Establishing the composition-structure-property relationships for amorphous materials is critical for many important natural and engineering processes, including the dissolution of highly complex volcanic glasses. In this investigation, we performed force field molecular dynamics (MD) simulations to generate detailed structural representations for ten natural CaO-MgO-$Al_2O_3$-$SiO_2$-$TiO_2$-FeO-$Fe_2O_3$-$Na_2O$-$K_2O$ glasses with compositions ranging from rhyolitic to basaltic. Based on the resulting atomic structural representations at 300 K, we have calculated the partial radial distribution functions, nearest interatomic distances and coordination number, which are consistent with the literature data on silicate-based glasses. Based on these structural attributes and classical bond valence models, we have introduced a novel structural descriptor, i.e., average metal-oxygen (M-O) bond strength parameter, which has captured the log dissolution rates of the ten glasses at both acidic and basic conditions (based on literature data) with $R^2$ values of ~0.80-0.92 based on linear regression. This structural descriptor is seen to outperform several other structural descriptors also derived from MD simulation results, including the average metal oxide dissociation energy, the average self-diffusion coefficient of all the atoms at their melting points, and the energy barrier of self-diffusion. Furthermore, we showed that the MD-derived descriptors generally exhibit better predictive performance than the degree of depolymerization parameter commonly used to describe glass and mineral reactivity. The results suggest that the structural descriptors derived from MD simulations, especially the average M-O bond strength parameter, are promising structural descriptors for connecting composition with dissolution rates of highly complex natural glasses.

# 1 Introduction

Silicate-based glasses are of significant interest to condensed matter physics, geology, glass science, materials chemistry, energy, medicine, advanced communication systems, metallurgical process, nuclear waste encapsulation and sustainable cement production [1, 2]. Natural volcanic glass produced due to the rapid cooling of magma is the most abundant silicate-based natural glass on the Earth, making up about 12% of the exposed continental crust surface [3] with an annual production rate of ~$10^9$ $m^3$ (mainly along with the 70,000 km oceanic ridge system) [4]. The abundance of natural volcanic glass, along with its chemical instability (prone to weathering), renders it important in the global and local geochemical cycle, including the cycling of global $CO_2$ [5] and the chemical composition of rivers, lakes, and soils [6]. Hence, understanding the dissolution behavior of natural volcanic glasses in different environmental conditions, including the impact of pH [7, 8], temperature [7, 8], solution chemistry [8], and bacterial activity [9], provides insight into potential future trends of these cycles.

A less studied aspect of natural volcanic glass dissolution is the inherent chemical variability of these glasses, which can vary significantly from one region to another and from one type of glass to another (e.g., from silicate-rich rhyolitic glasses to alkaline earth-rich basaltic glasses)[7].

This chemical variability impacts the dissolution rates of these natural volcanic glasses in both acidic and alkaline conditions [7, 10]. In particular, the impact of chemical variability on the reactivity (or dissolution behavior) of these volcanic glasses in alkaline conditions is also critical to the applications of volcanic ashes in concrete production. On the one hand, volcanic ash has been used as a natural pozzolan to partially replace ordinary Portland cement (OPC) in concrete [11], which is currently responsible for 8-9% of global anthropogenic $CO_2$ emissions [12]. On the other hand, volcanic ash has been used as a precursor to synthesize alkali-activated materials (AAMs) [13], which can exhibit a 40-80% reduction in $CO_2$ emissions compared with OPC [14]. Both applications exhibit great potential to lower the $CO_2$ footprint of the cement and concrete industry due to the natural abundance of volcanic ashes. The reactivity (or the rate of dissolution) of these volcanic ashes in concrete is critical to the development of concrete properties (e.g., strength). Therefore, it is essential to establish the relationship among composition, structure and properties (e.g., reactivity or dissolution behavior) for volcanic glasses.

The key to establishing the important composition-structure-properties relationship for these highly complex and amorphous materials is to obtain detailed structural information and develop reliable structural descriptors. One structural descriptor that has been used in the literature [15, 16] to connect glass or mineral composition to their dissolution rates is the extent of depolymerization parameter, i.e., the number of non-bridging oxygen (NBO) per network former T (NBO/T). NBO is defined as an oxygen atom connected to only one network former T (e.g., tetrahedrally coordinated Si and Al). Mineral dissolution studies have shown a positive correlation between NBO/T (calculated from mineral composition) and mineral dissolution rates [15]; however, the dissolution rate of alkaline earth silicate minerals can vary several orders of magnitude for minerals with the same level of NBO/T (e.g., the Si dissolution rate at pH 2 for orthosilicates (including $Mg_2SiO_4$, $Ca_2SiO_4$, $Mn_2SiO_4$, and $Fe_2SiO_4$) with NBO/T = 0 varies from ~$10^{-3}$ to ~$10^{-8}$ $m^{-2}s^{-1}$) [15, 17, 18]. For aluminosilicate glasses, some studies [16, 19] have reported a positive correlation between the NBO/T value of the glass and its reactivity in an alkaline environment, whereas other investigations have suggested that NBO/T is not always a reliable indicator of glass reactivity [18, 19].

Although NBO/T can be readily estimated from glass compositions following some simple rules [20], it is challenging to accurately estimate it for highly complex multi-components volcanic glasses (e.g., CaO-MgO-$Al_2O_3$-$SiO_2$-$TiO_2$-FeO-$Fe_2O_3$-$Na_2O$-$K_2O$ (CMASTFNK) glass system). Alternatively, simple empirical parameters calculated based on glass composition alone have also been used to describe the reactivity or dissolution of glasses [7, 21, 22]. One important example is the cement and concrete field, where empirical compositional parameters (e.g., the (CaO+MgO)/$SiO_2$ ratio for the European Standard for slag cement (1994)) are often used to describe the reactivity of supplementary cementitious materials (often rich in amorphous aluminosilicates) used in concrete production [21, 22]. However, the reliability of these empirical compositional parameters is often questionable, as illustrated by a recent study [22]. Hence, there is a general need to develop more reliable parameters to predict glass reactivity or dissolution rate from glass composition, and the development of structural descriptors containing detailed glass structural information is promising.

Although there are many experimental characterizations of volcanic glasses, most of existing investigations focused on the local structure of a single or several metal cations (e.g., coordination of Al, Ti or Fe) [23, 24, 25, 26, 27]. For these complex glasses, it is challenging to obtain a complete picture of their atomic structure (especially regarding its mid-range structure) using experiments alone  due to their amorphous nature [28]. For this purpose, molecular dynamics (MD) simulations have been increasingly used to generate detailed structural representations for different glass systems [29]. Based on MD simulation results and topology constraint theory, an important structural descriptor, i.e., the average number of constraints per atom in the glass network (an estimation of the network rigidity), has been developed and used to predict the dissolution rates of aluminosilicate glasses [30, 31], with better performance than NBO/T [18]. However, estimation of topology constraints for complex CMASTFNK glasses has not yet been reported in the literature as far as the authors are aware.

Furthermore, Force field MD-derived structural information has also been used by Lusvardi *et al.* to estimate the overall strength of the glass network ($F_{net}$), which was seen to correlate well with the leaching rate of bioactive fluorophospho-silicate glasses [32]. Recently, Lu *et al.* has used a slightly modified $F_{net}$ descriptor to accurately capture the initial dissolution rate of $ZrO_2$-containing soda-lime borosilicate glasses [33]. More recently, Gong and White used force field MD simulations to derive two structural descriptors for quaternary CaO-MgO-$Al_2O_3$-$SiO_2$ (CMAS) and ternary CaO-$Al_2O_3$-$SiO_2$ (CAS) glasses [34], i.e., (i) the average metal-oxygen dissociation energy (AMODE) parameter similar to the modified $F_{net}$ proposed by Lu *et al.* [33], and (ii) the average self-diffusion coefficient (ASDC) at melting. Both the AMODE and ASDC parameters have been shown to outperform the NBO/T parameter in predicting the relative reactivity of CMAS and CAS glasses in different alkaline environments [34]. These investigations have highlighted the importance of incorporating detailed structural information in developing reliable structural descriptors [30, 31, 32, 33, 34] (as opposed to those empirical parameters based on composition alone).

However, there is still a lack of reliable structural descriptors based on detailed structural information for predicting the dissolution rate of complex volcanic glasses. Although there are some MD investigations on the structure and properties of volcanic melt/glass [35, 36], most focus on high temperature/pressure properties, and none have used the MD-derived structural information to predict the dissolution rate of volcanic glasses as far as we are aware. Hence, in this study, we employed force field MD simulations to generate detailed atomic structural representations for ten volcanic CMASTFNK glasses with different compositions (ranging from rhyolitic to basaltic glasses). Based on analysis of resulting structural representations, several important structural attributes for each glass composition have been determined, including radial distribution functions (RDFs), the nearest interatomic M-O distances (M = Ca, Mg, Al, Si, Ti, Fe, Fe, Na, and K) and coordination numbers (CNs), which have been compared with the literature data on silicate-based glasses. Based on the structural analysis results and classical bond valence models, we have derived a new structural descriptor, i.e., average metal-oxygen bond strength parameter, and evaluated its performance in predicting the dissolution rate data in both acidic and alkaline conditions obtained from the literature on the same glasses, in comparison with the AMODE, ASDC at melting, the energy barrier of self-diffusion, as well as the commonly used NBO/T. This investigation represents a critical step forward in establishing the important composition-structure-dissolution relationships for highly complex volcanic glasses, which are essential to many natural and engineering processes.

# 2 Computational details

## 2.1 Glass compositions

Ten natural glasses with a range of chemical compositions (from rhyolitic to basaltic) have been selected from a high-quality experimental study [7] which has experimentally investigated the impact of glass composition on the dissolution rates in both acidic and alkaline conditions. The main chemical compositions and density of the ten glasses are summarized in Table 1, which shows that the glasses contain ~50-73 wt. % $SiO_2$, ~12-16 wt. % $Al_2O_3$, ~0-10 wt. %

CaO, ~0-6 wt. % MgO, ~0-3 wt. % TiO2, ~0-12 wt. % FeO, ~1-4 wt. % $Fe_2O_3$, ~2-6 wt. % $Na_2O$, and ~0-4 wt. % $K_2O$. Note that all the glasses also contain trace amounts of MnO (< ~0.3 wt. %) and $P_2O_5$ (< ~1 wt. %), which were not included in the subsequent MD simulations due to their relatively small quantities. Figure 1 shows the relationship between the $SiO_2$ content and the quantities of the different modifiers, which are seen to be highly correlated. Strong correlations are also observed among most network formers and network modifiers (see Fig. S1 of the Supplementary Material). Similar interconnections between individual oxide components have also been observed in blast-furnace slags in previous investigations [34, 37]. Hence, the different dissolution rates of these highly complex glasses should not be simply attributed to the relative quantity of one individual oxide component. Instead, it is necessary to derive physics-based structural descriptors for a more accurate description of the composition-structure-properties relationship for these highly complex and important glasses.

Table 1. The chemical composition of the main oxides (in weight percentage) and the density of the ten glasses. Also given in the table are the log dissolution rate data (based on Si atom and normalized by BET surface area) collected on these glasses at a pH of 4 and 10.6 under far-from-equilibrium conditions. All the data are obtained from reference [7].

| Glass ID | Main oxide composition (wt. %) | | | | | | | | | Volcanic name | Density ($g/cm^3$) | Log dissolution rate, $\log r_{+\,BET}$ ($mol/m^2/s$) | |
|---|---|---|---|---|---|---|---|---|---|---|---|---|---|
| | $SiO_2$ | $Al_2O_3$ | CaO | MgO | $TiO_2$ | FeO | $Fe_2O_3$ | $Na_2O$ | $K_2O$ | | | pH = 10.6 | pH = 4 |
| 1BT | 72.62 | 12.48 | 0.48 | 0.09 | 0.08 | 0.11 | 0.78 | 3.79 | 4.38 | rhyolite | 2.31 | -10.30 | -11.53 |
| 2O62 | 70.64 | 13.00 | 0.97 | 0.02 | 0.24 | 1.18 | 2.40 | 5.45 | 3.41 | rhyolite | 2.36 | -9.78 | -10.99 |
| 5A75 | 69.28 | 12.42 | 2.81 | 0.97 | 0.90 | 2.09 | 2.48 | 3.74 | 2.21 | rhyolite | 2.45 | -10.06 | -11.17 |
| 6H3B | 66.01 | 14.65 | 3.21 | 0.39 | 0.42 | 3.81 | 2.14 | 4.72 | 2.07 | dacite | 2.51 | -9.65 | -10.72 |
| 7HZ0 | 62.80 | 15.34 | 4.57 | 1.40 | 0.96 | 4.53 | 2.64 | 4.35 | 1.55 | dacite | 2.59 | -9.35 | -10.86 |
| 9H20 | 54.81 | 14.35 | 6.65 | 2.81 | 1.98 | 7.48 | 4.38 | 4.01 | 1.27 | b-andesite | 2.79 | -8.36 | -10.06 |
| 12GR | 50.77 | 13.55 | 9.87 | 5.54 | 2.53 | 10.26 | 2.57 | 2.77 | 0.52 | basalt | 2.99 | -8.47 | -9.83 |
| 13HEI | 50.48 | 16.16 | 6.94 | 2.54 | 2.24 | 7.77 | 3.78 | 5.84 | 1.60 | mugearite | 2.81 | -8.67 | -9.41 |
| 15KRA | 49.78 | 13.44 | 10.23 | 5.73 | 2.01 | 11.56 | 3.06 | 2.37 | 0.32 | basalt | 3.04 | -8.97 | -9.77 |
| 16KAT | 47.07 | 12.65 | 9.11 | 4.79 | 4.50 | 11.08 | 4.13 | 2.79 | 0.77 | basalt | 3.04 | -9.21 | -10.37 |

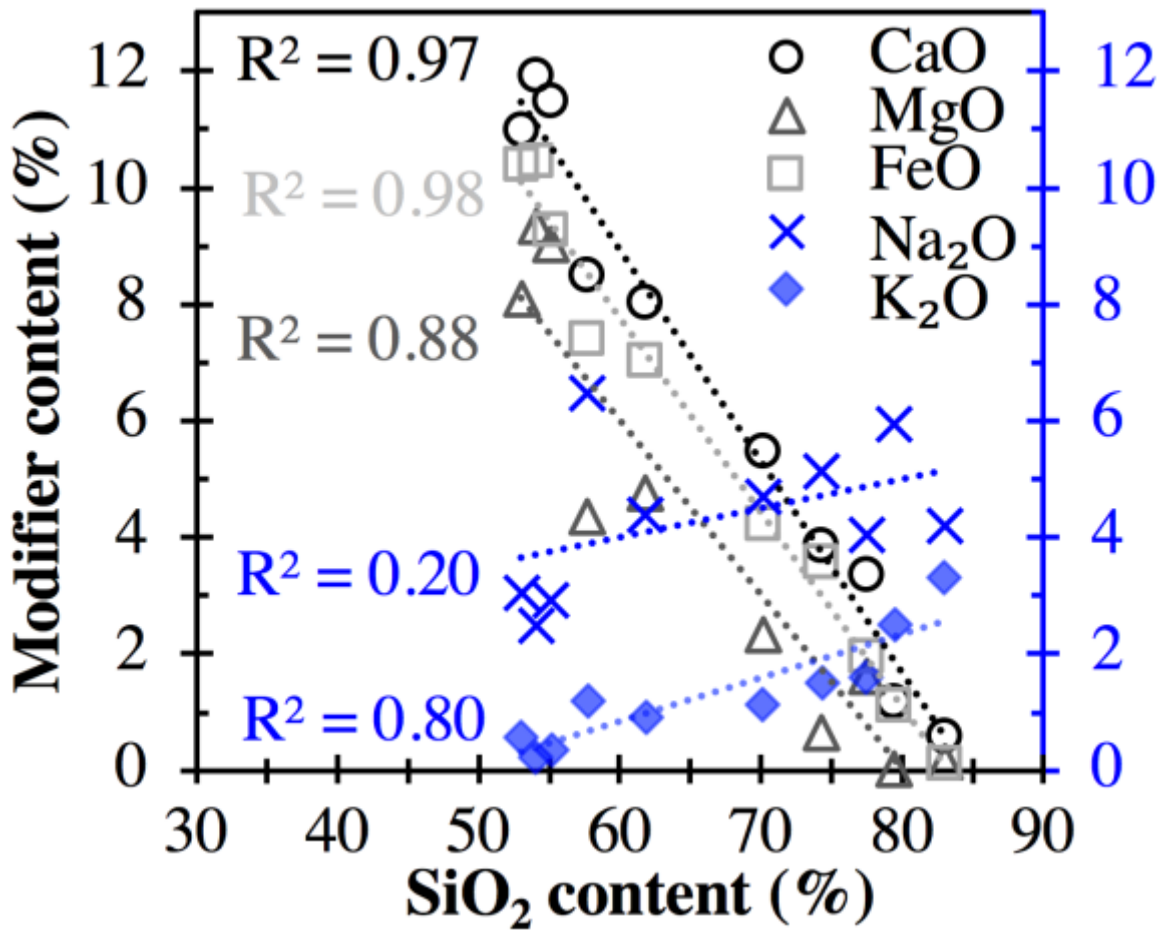

Figure 1. Correlation between $SiO_2$ content and individual modifier content (i.e., CaO, MgO, FeO, $Na_2O$ and $K_2O$) in molar percentage for the glasses in Table 1. $R^2$ values for linear regression are given in the figure.

## 2.2 Force fields

Force field MD simulations following a commonly used "melt-and-quench" process have been used in this investigation to generate amorphous structural representations for the volcanic glasses in Table 1. We employed the force field that was developed by Pedone *et al.* [38] to cover complex silicate crystals, melts and glasses in the compositional space of CaO-MgO-$Al_2O_3$-$SiO_2$-$TiO_2$-FeO-$Fe_2O_3$-$Na_2O$-$K_2O$ for all the MD simulations. The Pedone force field is described by Equation (1) [38].

$$U_{ij}(r_{ij}) = \frac{z_i z_j}{r_{ij}} + D_{ij}\left[\left\{1 - e^{-a_{ij}(r_{ij}-r_0)}\right\}^2 - 1\right] + \frac{C_{ij}}{{r_{ij}}^{12}} \quad (1)$$

where the three terms represent (i) a long-range Coulomb interaction, (ii) a short-range Morse function, and (iii) a repulsive contribution to avoid collapse at high temperature and pressure, respectively. $D_{ij}$, $a_{ij}$, $r_0$, and $C_{ij}$ are empirical parameters derived by fitting data collected on various silicate crystals [38]. Due to the difficulty of taking into account both $^{II}Fe$ and $^{III}Fe$ in the simulations at the same time, we have performed the MD simulations on each glass composition in Table 1 twice by assuming all iron to be either in FeO or $Fe_2O_3$. This means that MD simulations will be performed using two sets of force field parameters, denoted as Pedone_Fe2, and Pedone_Fe3, which has been summarized in Table S1 of the Supplementary Material. This also allows the impact of $^{II}Fe$ and $^{III}Fe$ on the glass structures to be explored.

## 2.3 Generation of glass structures

Tables 2 and 3 give a summary of the elemental compositions of each glass simulation box (a cubic cell with around 6000 atoms), assuming that iron is 100% FeO and $Fe_2O_3$, respectively. For each structure, the atoms were initially randomly placed in the simulation box, and the density of the unit cell was set at a value estimated for each glass at a temperature of 5000 K. The value was estimated using a similar method adopted in previous investigations [34, 39], with detailed calculation given in Section 3 of the Supplementary Material. The structure was firstly equilibrated at 5000 K for 500 ps to ensure the loss of the memory of the initial configuration. It was then progressively quenched from 5000 to 3000 (over 1000 ps), 3000 to 2000 (over 500 ps), 2000 to 1000 (over 500 ps), and 1000 to 300 (over 500 ps) K, using a canonical *NVT* ensemble with the Nosé Hoover thermostat and a time step of 1 fs. At each temperature (i.e., 3000, 2000, 1000, and 300 K), the density of the cell was adjusted to either numerically estimated (for high-temperature density) or experimental values (density values at 300 K are obtained from the literature [7], as given in Table 1) before being equilibrated for 1000 ps using the same canonical *NVT* ensemble and time step. Similar procedures using *NVT* ensemble have been adopted previously to generate reasonable amorphous structural representations for CMAS and CAS glasses [34].

Based on the MD trajectories of the final 500 ps of the *NVT* equilibration step at 300 K (containing 500 structural snapshots), detailed structural analysis has been performed for all glasses, including the calculation of radial distribution functions (calculation details are given in Section 4 of the Supplementary Material), the nearest interatomic distances, and CNs of different metal cations in their first coordination shell. It has been shown previously for CMAS and CAS glasses that the interatomic distances and coordination numbers exhibit limited variation from one production run to another for the same glass composition. Nevertheless, we have repeated the simulations three times for all the glasses in Table 3 using the Pedone_Fe3 force field to evaluate the reproducibility of the results. All the simulations were carried out using the ATK-Forcefield module implemented in the QuantumATK software package [40, 41].

Table 2. The number of atoms in each simulation box for each glass in Table 1 assuming all Fe atoms are in FeO (i.e., $^{II}$Fe).

| Glass ID | Si | Al | Ca | Mg | Ti | $^{II}$Fe | Na | K | O | Total |
|---|---|---|---|---|---|---|---|---|---|---|
| 1BT_Fe2 | 1574 | 318 | 11 | 3 | 1 | 14 | 160 | 122 | 3796 | 5999 |
| 2O62_Fe2 | 1510 | 326 | 22 | 1 | 4 | 60 | 226 | 94 | 3760 | 6003 |
| 5A75_Fe2 | 1497 | 316 | 65 | 31 | 15 | 78 | 156 | 60 | 3780 | 5998 |
| 6H3B_Fe2 | 1428 | 372 | 74 | 13 | 7 | 104 | 198 | 58 | 3747 | 6001 |
| 7HZ0_Fe2 | 1361 | 392 | 106 | 46 | 16 | 124 | 182 | 42 | 3730 | 5999 |
| 9H20_Fe2 | 1232 | 380 | 160 | 95 | 33 | 214 | 174 | 36 | 3674 | 5998 |
| 12GR_Fe2 | 1147 | 360 | 239 | 188 | 43 | 236 | 122 | 14 | 3651 | 6000 |
| 13HEI_Fe2 | 1142 | 430 | 168 | 86 | 38 | 210 | 256 | 46 | 3620 | 5996 |
| 15KRA_Fe2 | 1132 | 360 | 249 | 195 | 34 | 272 | 104 | 10 | 3645 | 6001 |
| 16KAT_Fe2 | 1100 | 348 | 228 | 168 | 79 | 288 | 126 | 22 | 3638 | 5997 |

Table 3. The number of atoms in each simulation box for each glass in Table 1 assuming all Fe atoms are in $Fe_2O_3$ (i.e., $^{III}$Fe).

| Glass ID | Si | Al | Ca | Mg | Ti | $^{III}$Fe | Na | K | O | Total |
|---|---|---|---|---|---|---|---|---|---|---|
| 1BT_Fe3 | 1574 | 318 | 11 | 3 | 1 | 14 | 160 | 122 | 3803 | 6006 |
| 2O62_Fe3 | 1510 | 326 | 22 | 1 | 4 | 60 | 226 | 94 | 3790 | 6033 |
| 5A75_Fe3 | 1497 | 316 | 65 | 31 | 15 | 78 | 156 | 60 | 3819 | 6037 |
| 6H3B_Fe3 | 1428 | 372 | 74 | 13 | 7 | 104 | 198 | 58 | 3799 | 6053 |
| 7HZ0_Fe3 | 1361 | 392 | 106 | 46 | 16 | 124 | 182 | 42 | 3792 | 6061 |
| 9H20_Fe3 | 1232 | 380 | 160 | 95 | 33 | 214 | 174 | 36 | 3781 | 6105 |
| 12GR_Fe3 | 1147 | 360 | 239 | 188 | 43 | 236 | 122 | 14 | 3769 | 6118 |
| 13HEI_Fe3 | 1142 | 430 | 168 | 86 | 38 | 210 | 256 | 46 | 3725 | 6101 |
| 15KRA_Fe3 | 1132 | 360 | 249 | 195 | 34 | 272 | 104 | 10 | 3781 | 6137 |
| 16KAT_Fe3 | 1100 | 348 | 228 | 168 | 79 | 288 | 126 | 22 | 3782 | 6141 |

# 3 Results & Discussion

## 3.1 Local atomic structure

### 3.1.1 Nearest interatomic distances

Figure 2a shows a typical atomic structural representation obtained for the 13HEI_Fe3 glass composition in Table 2 using the Pedone_Fe3 force field parameters (Table S1 of the Supplementary Material), which is seen to be a highly disordered aluminosilicate network structure. The corresponding RDF for this structure is given in Figure 2b, which shows evident local (< ~3 Å) and medium-range (~3-10 Å) structural ordering but lack of long-range ordering above ~10 Å, in agreement with X-ray and neutron total scattering data reported on silicate-based glasses [34, 37, 39, 42, 43, 44, 45, 46, 47, 48] The assignments of the peaks can be facilitated by calculating the contribution from individual atom-atom pair, i.e., partial RDFs, as shown in Figure 2c, for the different M-O pairs (M = Ca, Mg, Al, Si, Ti, $^{III}$Fe, Na, and K). From the peak position of the partial RDFs in Figure 2c, the nearest interatomic M-O distances have been determined, which are summarized in Table S2 of the Supplementary Material for all the glass compositions investigated.

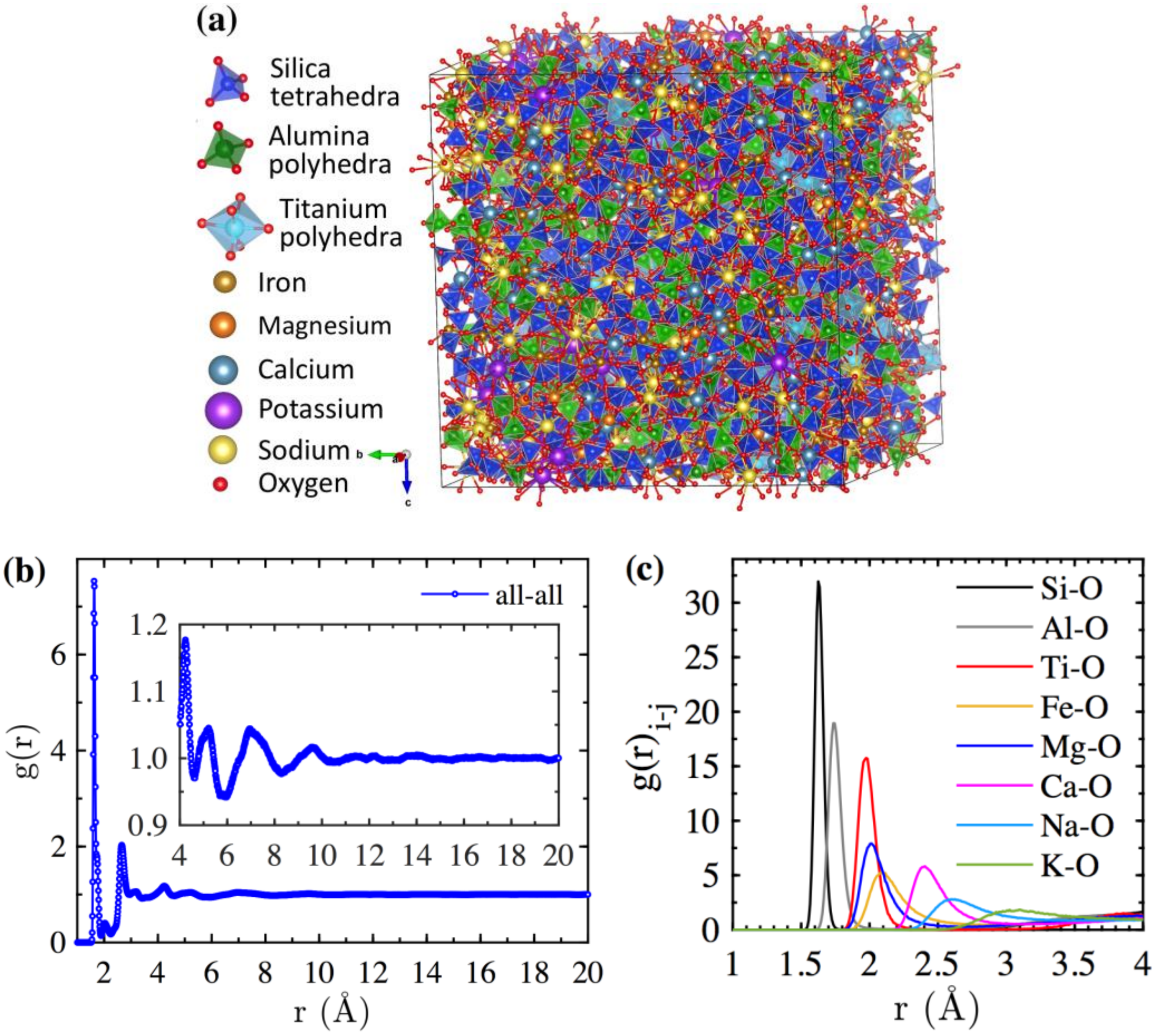

Figure 2. (a) A typical atomic structural representation (at 300 K) obtained for the 13HEI_Fe3 glass composition in Table 3, and (b) the corresponding radial distribution function (RDF). (c) shows the partial RDF for different M-O pairs (M = Ca, Mg, Al, Si, Ti, $^{III}$Fe, Na, and K) for the same 13HEI_Fe3 glass structure. The calculation of RDF is based on the last 500 ps of MD trajectories during the *NVT* equilibration step at 300 K, and the calculation details are given in Section 4 of the Supplementary Material.

The average nearest M-O bond distance values are given in Table 4 (with one standard deviation given in the brackets) and compared with those reported in the literature for silicate-based glasses (since data on similar volcanic glasses are rare). The relatively small standard deviation for M-O distances in Table 4 (mostly less than 0.03 Å) illustrates that the variation in glass composition has a limited impact on the nearest interatomic distances, which is consistent with previous MD simulations on CAS and CMAS glasses [34]. The comparison in Table 4 shows that the predicted interatomic distances are generally consistent with the reported experimental and simulation data on silicate-based glasses in the literature. For example, X-ray absorption near edge structure (XANES) data on rhyolitic glasses suggested an $^{III}$Fe-O distance of 1.85±0.02 Å [23], while a study on $Na_2O$-$Fe_2O_3$-$SiO_2$ glasses based on neutron diffraction and empirical potential structure refinement (EPSR) modeling give an $^{III}$Fe-O distance of 1.87 Å, which are in an excellent agreement with the values obtained here (~1.86 Å). For $^{II}$Fe-O bond distance, values of 2.01-2.02 [48], 2.04-2.08 [47], and 2.04 Å [49] have been reported for CaO-$FeO_x$-$SiO_2$ (CFS) glasses based on either X-ray/neutron scattering data or MD simulations, which are also well captured by our MD simulations (i.e., 2.03 Å). The fact that the $^{II}$Fe-O distances are larger than $^{III}$Fe-O distances but similar to Mg-O distances (~2.00 Å) is consistent with the general view that $^{II}$Fe acts more like a network modifier (similar to $Mg^{2+}$ and $Ca^{2+}$) while $^{III}$Fe acts more like a network former (similar to $Al^{3+}$).

Furthermore, the force field is seen to slightly underpredict the Ti-O, Al-O and O-O distances, however, the differences are within 2% of the reported values. We note that there is limited experimental and simulation data collected on complex volcanic glasses for a direct comparison with our simulation results. Given that the glass composition, e.g., the type and amount of modifier content [50] and the amount of $TiO_2$ content [51], has been shown to have a visible impact on Ti-O bond distance in binary and ternary silicate glasses, the underprediction for Ti-O distances (compared with the literature data) could be partially attributed to the differences in the glass composition. Finally, the impact of $^{II}$Fe versus $^{III}$Fe on the M-O bond distances of other metal cations (Table 4) is seen to be limited.

Table 4. Summary of the nearest interatomic distances (averaged over the ten glass compositions shown in Tables 2 and 3), in comparison with the literature experiment and simulation data on silicate-based glasses. The values in the brackets are one standard deviation based on all the ten compositions, and the raw data are given in Table S2 of the Supplementary Material. Literature data (from both experiments and simulation) on different types of silicate glasses (rhyolitic glasses [23], $CaO$-$MgO$-$Al_2O_3$-$SiO_2$ (CMAS) [34, 37, 39], $MgO$-$Al_2O_3$-$TiO_2$-$SiO_2$ (MATS) [52], $CaO$-$MgO$-$K_2O$-$SiO_2$ (CMKS) [45], $CaO$-$Al_2O_3$-$SiO_2$ (CAS) [42, 53, 54, 55, 56], $MgO$-$Al_2O_3$-$SiO_2$ (MAS) [44], $FeO$-$Al_2O_3$-$SiO_2$ (FAS) [47, 48], $CaO$-$Na_2O$-$SiO_2$ (CNS) [57, 58], $CaO$-$FeO$-$SiO_2$ (CFS) [49], $CaO$-$TiO_2$-$SiO_2$ (CTS) [50], $Na_2O$-$TiO_2$-$SiO_2$ (NTS) [50], $Na_2O$-$FeO$-$SiO_2$ (NFS) [59], $K_2O$-$TiO_2$-$SiO_2$ (KTS) [50], $Na_2O$-$Al_2O_3$-$SiO_2$ (NAS) [58, 60, 61], $K_2O$-$SiO_2$ (KS) [43, 62, 63, 64, 65], $Na_2O$-$SiO_2$ (NS) [57, 58, 62, 63, 64], and $TiO_2$-$SiO_2$ (TS) [51]) are given for comparisons. EXAFS: Extended X-ray absorption fine structure; XANES: X-ray absorption near edge structure; NMR: Nuclear magnetic resonance; RMC: Reverse Monte Carlo; DFT: Density functional theory; MD: Molecular dynamics; FF MD: Force field molecular dynamics; EPSR: Empirical potential structure refinement.

| Atom-atom pair | Nearest interatomic distance (Å) | | | |
|---|---|---|---|---|
| | Pedone_Fe2 | Pedone_Fe3 | Experimental data in the literature | Simulation data in the literature |
| Si-O | 1.60 (0.00) | 1.60 (0.01) | 1.61-1.66, X-ray and neutron diffraction [39, 43, 44, 47, 48, 53, 54, 55, 56]; 1.60, EXAFS [57]; | 1.62-1.63, FF MD [34, 39, 49]; 1.62, RMC [44, 52]; 1.61-1.64, *ab initio* MD [61] |
| Al-O | 1.72 (0.00) | 1.73 (0.00) | ~1.74-1.77, X-ray/neutron diffraction [39, 44, 53, 54, 55, 56] | 1.74-1.76 FF MD [34, 39]; 1.73-1.77, RMC [44, 52]; 1.75-1.77, *ab initio* MD [61] |
| Ti-O | 1.80 (0.01) | 1.79 (0.01) | 1.85-1.89, XANES [50]; 1.86, neutron diffraction [52] | 1.83-1.93, *ab initio* MD [51]; 1.85, RMC [52] |
| $^{II}$Fe-O | 2.03 (0.02) | -- | 2.01-2.08, X-ray/neutron scattering [47, 48]; | 2.04, FF MD [49] |
| $^{III}$Fe-O | -- | 1.86 (0.01) | 1.85±0.02, XANES [23] | 1.87, EPSR [59] |
| Mg-O | 1.99 (0.02) | 2.00 (0.02) | ~2.00, X-ray/neutron diffraction [37, 39, 44, 45] | 2.03-2.04, FF MD [34, 39]; 2.02, DFT [39]; 2.00-2.05, RMC [44, 52]; |
| Ca-O | 2.36 (0.01) | 2.37 (0.01) | ~2.25-2.39, X-ray and/or neutron diffraction [37, 39, 42, 45, 48, 54, 55, 56]; | 2.40-2.45, FF MD [34, 39, 49]; 2.35, DFT [39] |
| Na-O | 2.38 (0.02) | 2.39 (0.02) | 2.3-2.43, EXAFS [57]; 2.34-2.36, X-ray/neutron diffraction [62]; 2.56-2.62, EXAFS and XANES [60]; 2.45-2.6, NMR [58] | 2.30, EPSR [59]; 2.36-2.37, *ab initio* MD [61]; 2.30-39, FF MD, NS/KS [63, 64] |
| K-O | 2.75 (0.03) | 2.72 (0.07) | 2.6-2.7, X-ray/neutron scattering [43, 45, 62]; | 2.67-2.81, FF MD [34, 63, 64, 65]; |
| O-O | 2.61 (0.01) | 2.61 (0.01) | 2.54-2.72, X-ray/neutron scattering [39, 44, 45, 47, 48, 56] | 2.66, DFT[39]; 2.70-2.72, *ab initio* MD [61]; 2.60, FF MD [49] |

### 3.1.2 Coordination chemistry

The evolution of oxygen coordination number (CN) with increasing cutoff distance for the different metal cations M (M = Si, Al, Ti, $^{III}$Fe, Mg, Ca, Na and K) in a typical glass composition (i.e., 13HEI_Fe3) are compared in Figure 3, which clearly shows that Si, Al and Ti atoms have relatively well-defined first coordination shells since a plateau of CN has reached above ~2.2 Å. In contrast, the CNs of the other metal cations are highly sensitive to the selected cutoff distance, with the following order of sensitivity: K > Na > Ca > Mg. This order appears to be associated with the field strength of the metal cations (defined as $Z/d^2$, where $Z$ is the cation charge and $d$ is the cation-oxygen distance), with a higher field strength giving a better defined first coordination shell (and hence being less sensitive to the selection of cutoff distance). To estimate the oxygen CNs within the first coordination shell of each metal cation, it is necessary to define the corresponding cutoff distance. Here, the cutoff distance for each M-O pair is determined from the first minima of the corresponding partial RDF, which is a common approach used in the glass modeling literature [34, 39, 49, 53, 66] (see Fig. S2 and Table S3 in the Supplementary Material for a summary of the cutoff distances for all the simulated glass compositions). Due to the relatively light impact of glass composition on the cutoff distances (Table S3), we used the same cutoff distance for each M-O pair (averaged over ten glass compositions) for ease of comparison. The adopted cutoff distances are 2.1, 2.4, 2.5, 2.6, 2.8, 2.7, 3.1, 3.4 and 3.9 Å for Si-O, Al-O, Ti-O, $^{II}$Fe-O, $^{III}$Fe-O, Mg-O, Ca-O, Na-O and K-O, respectively.

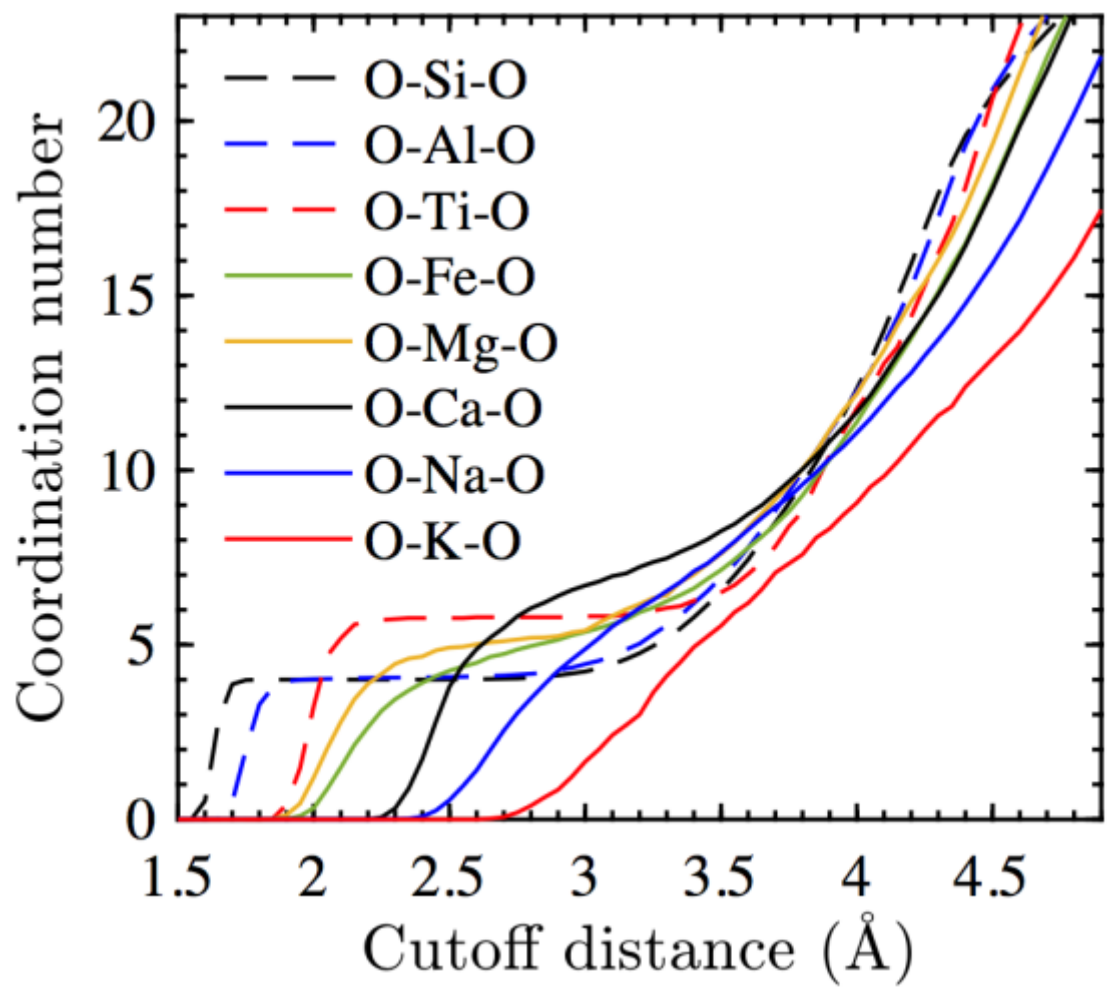

Figure 3. Evolution of coordination number (i.e., number of oxygen atoms surrounding Si, Al, Ti, Fe, Mg, Ca, Na and K atoms) as a function of cutoff distance for the Si/Al/Ti/$^{III}$Fe/Mg/Ca/Na/K-O pairs in the 13HEI_Fe3 glass in Table 2.

The distributions of oxygen CN for all the metal cations M in a typical glass composition (i.e., 13HEI) along with the impact of $^{II}$Fe and $^{III}$Fe (i.e., 13HEI_Fe2 in Table 2 versus 13HEI_Fe3 in Table 3) are shown in Figure 4, while the average CNs are summarized in Table 5 in comparison with literature data on silicate-based glasses. A complete summary of the average CNs for all the glass compositions investigated here is given in Table S4 of the Supplementary Material. It is clear from Figure 4 and Table 5 that Si atoms are almost in 100% IV-fold coordination (Figure 4a) and that Al atoms are also predominantly in IV-fold coordination along with a few percent of V-fold and negligible VI-fold coordination (Figure 4b), which are consistent with literature data on aluminosilicate glasses (see Table 5). Furthermore, $^{II}$Fe cation is seen to be dominated by V-fold coordination state with considerable IV- and VI-fold coordination, whereas $^{III}$Fe cation is dominated by IV-fold coordination with sizeable V-fold (Figure 4d). These results are also consistent with literature data on Fe coordination in aluminosilicate glasses (see Table 5 for a summary), including volcanic glasses with compositions similar to those shown in Table 1. For example, XANES studies on rhyolitic glasses [6] and basaltic/felsic glasses [41] have shown that $^{II}Fe$ and $^{III}Fe$ atoms are mainly in V-fold and IV-fold coordination, respectively. Another XANES investigation on rhyolitic glasses [42] has estimated the average CN of $^{III}Fe$ cation to be ~4.6, which is close to the average CN values of $^{III}Fe$ in Figure 4d (i.e., ~4.4).

The Ti atom in the 13HEI glass (Figure 4c) is seen to mainly adopt IV- and V-fold coordination with a small amount of VI-fold. Examination of the literature data on Ti coordination in silicate glasses (see Table 5; brief reviews of Ti coordination chemistry in silicate glasses have been given in references [67, 68]) reveals that the dominant Ti coordination varies considerably from IV-fold to VI-fold depending on the type and concentration of the alkali or alkaline earth cations, the $TiO_2$ content, and the degree of polymerization [51, 67, 68, 69]. According to a XANES study on natural volcanic glasses [25] with compositions comparable to those shown in Table 1, V-fold coordination is the dominant Ti coordination in all the volcanic glasses studied (from rhyolitic to basaltic composition), and that the average CN is in the range of 4.4-5.5, which is higher those obtained here (4.3-5.0, Table 5). This underestimation of Ti CN suggests that refinement of the Pedone force field is needed to more accurately capture Ti coordination in these highly complex natural volcanic glasses, which is beyond the scope of the current investigation but worth exploring in the future. Nevertheless, comparison of the simulated Ti CN with the experimental data [25] (see Fig. S3 in the Supplementary Material) showed that the overall trend of evolution of Ti CN as a function of NBO/T has been captured by our simulations.

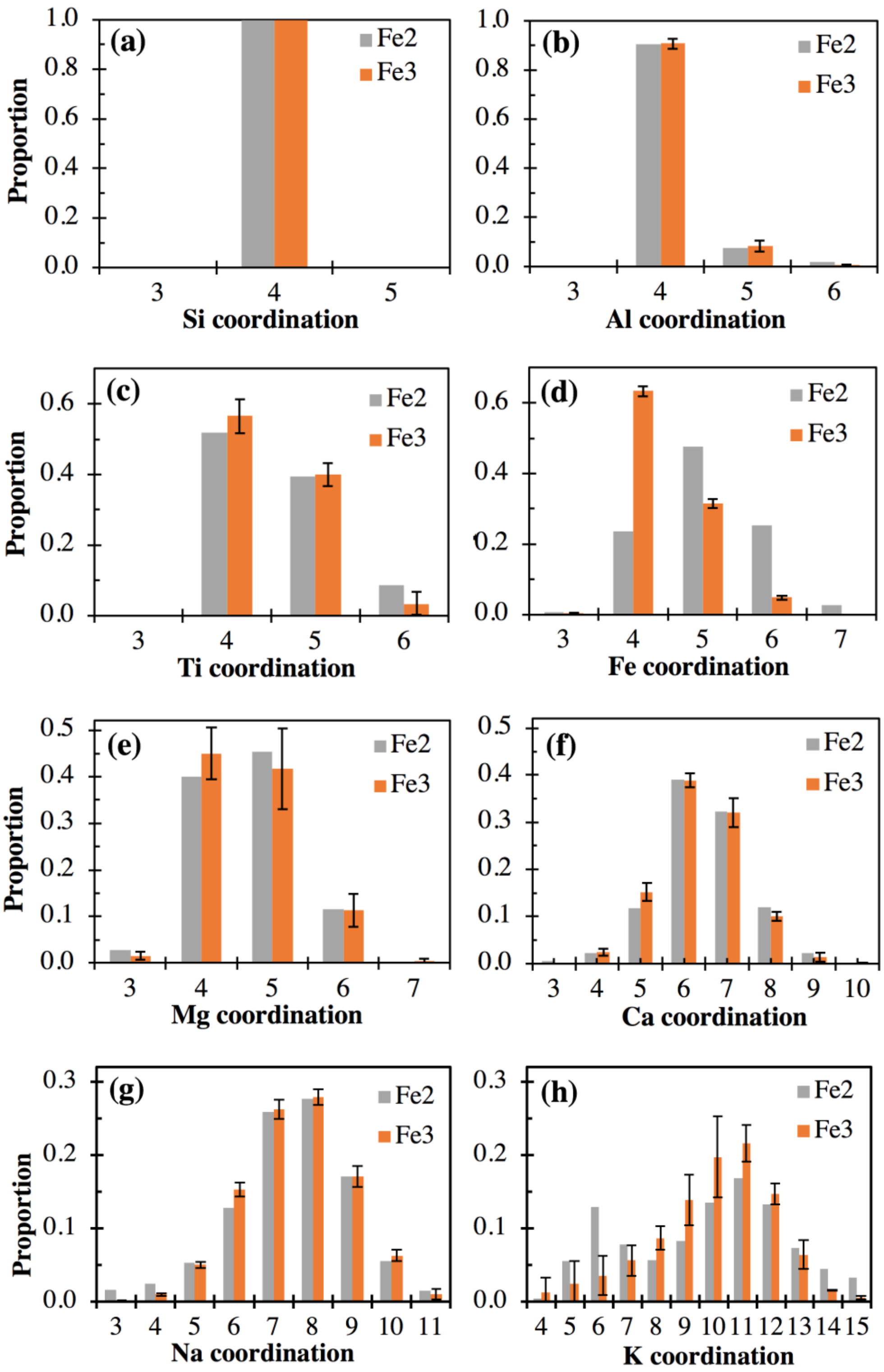

Figure 4. Distributions of coordination number (CN) for (a) Si, (b) Al, (c) Ti, (d) Fe, (e) Mg, (f) Mg (g) Na and (h) K atoms in typical glass structures (i.e., 13HEI_Fe2 and 13HEI_Fe3). The error bars are one standard deviation based on three independent MD production runs.

Figure 4e shows that Mg cation in the 13HEI glass (a mugearite glass) is dominated by IV-, and V-fold coordination, whereas Ca atom is seen to adopts mainly VI- and VII-fold coordination with sizable V- and VIII-fold coordination (Figure 4f). The average CN of Mg and Ca cations in all the glasses investigated here are seen to be around 4.2-4.9 and 6-6.7, respectively (Table 5). These results are generally consistent with previous investigations on aluminosilicate glasses, where V-fold Mg [44, 70, 71] and VI- and VII-fold Ca [55, 66, 72, 73, 74] are often reported as the dominant species. Nevertheless, previous $^{25}Mg$ nuclear magnetic resonance (NMR) studies on MAS and CMAS glasses show Mg is mainly in 6-coordination [73, 75], and XANES results on haplobasaltic glasses ($SiO_2$-$Al_2O_3$-CaO-MgO-$Na_2O$-FeO) suggests VII- and/or VIII-fold as the dominant coordination states for Ca [76]. In addition to the differences in chemical compositions, the discrepancy between different experimental results can be partially attributed to the sensitivity of different experimental techniques to the local bonding environments, as has been discussed in reference [77] for XANES and NMR.

Figure 4g and 5h show the CN distribution of Na and K cations in the 13HEI glass, respectively, where the distributions are seen to be much wider than those of the other metal cations (Figure 4a-f). The average CNs of the two alkali metal cations, especially the K cation (~8.4-11.6), are seen to be significantly higher than the other metal cations shown in Table 5. This average CN appears to be governed by the field strength of the metal cation, as illustrated in Figure 5, where the average CN of a metal cation is seen to be inversely correlated with its field strength. A similar inverse correlation is also observed in Figure 5 between the field strength and the number of coordination states adopted by the metal cation. Comparison of the average Na and K CN values with those reported in the literature data on silicate-based glasses (Table 5) shows that the Pedone force field generally gives slightly higher CN values for the volcanic glasses studied here.

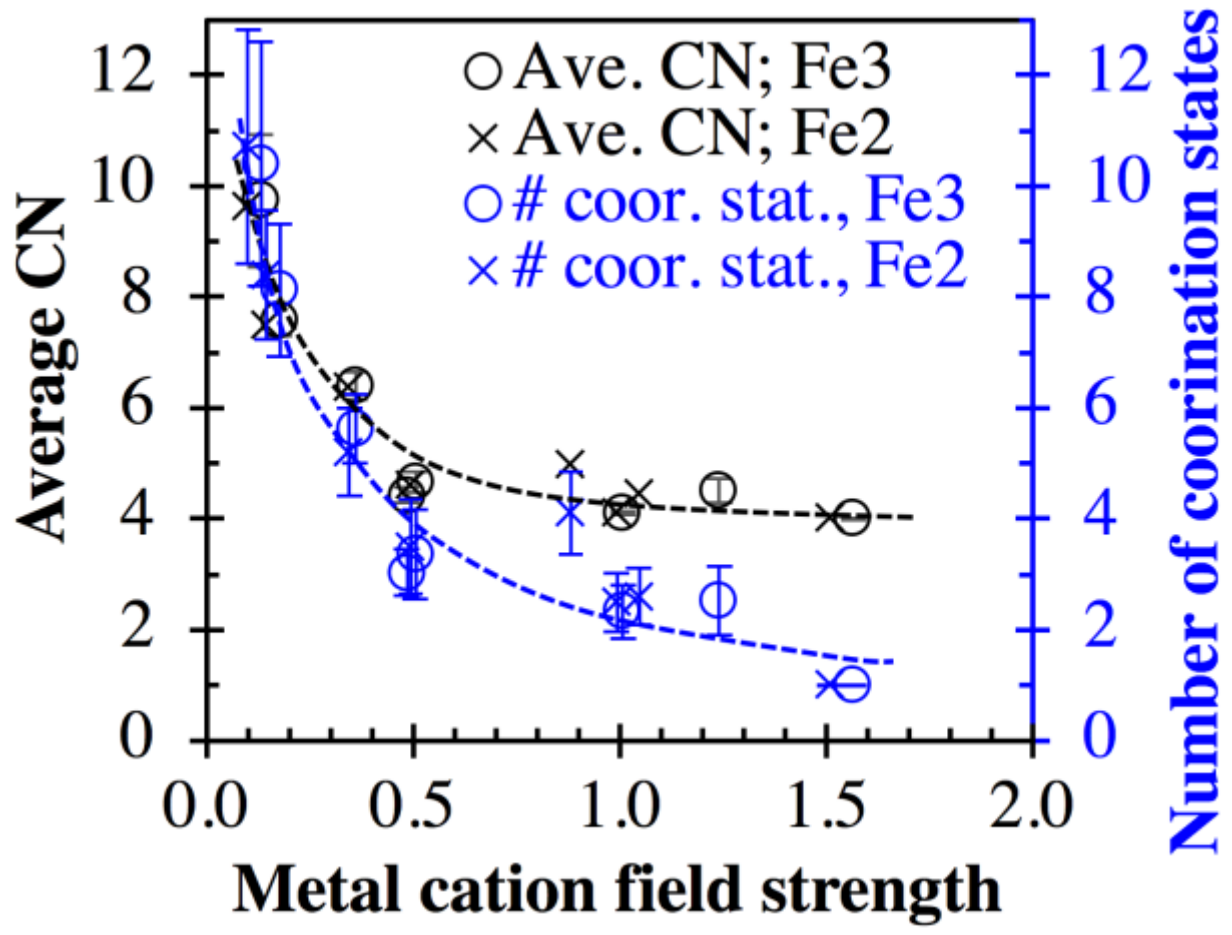

Figure 5. Correlation between metal cation field strength (defined as $Z/d^2$, where $Z$ is the cation charge and $d$ is the cation-oxygen distance) and (i) the average CN of the metal cation and (ii) the number of coordinate states adopted by the metal cation (only coordination states exceeding 1% have been considered). The error bars in the figure are one standard deviation based on results from three independent MD production runs. The dashed lines in the figure are given to guide the eye.

However, we note that the literature data for Na and K cations (Tables 4 and 5) are reported mostly for simpler binary or tertiary silicate glasses, and there is still a lack of experimental and simulation data on room temperature CNs and interatomic distances for Na and K cations in these highly complex volcanic glasses for a more direct comparison. This dramatic difference in compositions may have contributed to the discrepancies in CN observed between the complex volcanic glasses studied here and simpler binary or ternary glasses. In addition, several other factors may have also contributed to these discrepancies. This includes (i) the use of different cutoff distances for the calculation of CNs in different MD simulation studies (e.g., [63] and [64] obtained a cutoff distance of 3.8 Å for K-O pair based on the first minimum of the corresponding partial RDFs while [65] obtained a cutoff value of 3.0 Å based on the maximum position of the $1^{st}$ order derivative of the K-O pair interaction potential), (ii) the use of different type of force fields and MD production procedures (e.g., quench rate and type of canonical ensemble), and (iii) the differences between the cutoff distances used here and those probed by specific experimental techniques (i.e., different experimental techniques have varied sensitivity to the local bond environment, which influence distances probed [77]).

In summary, we show in this section that reasonable amorphous structural representations have been generated for volcanic glasses in Table 1 using the "melt-and-quench" method in MD simulations and the Pedone force field [38]. The resulting atomic structural features (i.e., the nearest interatomic distances and CN) are seen to be generally consistent with the literature data reported on volcanic glasses or simpler silicate-based glasses. Comparing the results for different volcanic glasses reveals that glass composition has a noticeable impact on CNs (especially for metal cations with low field strength, e.g., K) but negligible effect on the nearest interatomic distances. Finally, the impact of $^{II}$Fe and $^{III}$Fe on the CN distribution (Figure 4) and the average CN (Table 5) of other metal cations is relatively mild.

Table 5. Summary of oxygen coordination numbers (CNs) for different metal cations in different types of silicate glasses (e.g., rhyolitic glasses [23, 24, 25, 27], andesite glasses [24], basaltic/felsic glasses [26], CMAS [66, 73], and NCAS [78], MATS [52], CAS [42, 55, 72, 79], MAS [44, 70], FCS [47], KTS [50, 69], NTS [50, 69], NAS [61], CTS [50], NFS [59], CFS [49], MS [46, 75], NS [62, 63, 64], KS [43, 62, 63, 64, 65], and TS [51]) from the literature data.

| Atom pairs | Current study | | Coordination number (CN) | |
|---|---|---|---|---|
| | PedoneFe2 | PedoneFe3 | Literature data from experiments | Literature data from simulations |
| Si-O | 4.00 | 4.00 | 4, NMR [73];<br>4.04, neutron diffraction [55] | 4, RMC refinement [44], FF MD [66], and EPSR [59];<br>4-4.02 *ab initio* MD [61] |
| Al-O | 4.07-4.16 | 4.06-4.18 | 4.00-4.02, NMR [24]; 4, NMR [73]; 4.19 NMR [52];<br>4.1-4.20 [55, 79] | 4.1, FF MD [66];<br>4.08-4.16, RMC refinement [44, 52]; 4.08-4.36, *ab initio* MD [61] |
| Ti-O | 4.25-4.63 | 4.27-4.99 | 4.8-5.3, XANES [50];<br>4.4-5.5, XANES [25];<br>4.6-5.3, XANES [69];<br>5.4, neutron diffraction [52] | 4.1-5.6, *ab initio* MD [51];<br>5.07, RMC [52] |
| $Fe^{III}$-O | | 4.24-4.69 | Mainly 4-fold, XANES [23, 26];<br>4.6, XANES [27] | 4.4, EPSR [59] |
| $Fe^{II}$-O | 4.79-5.15 | | Mainly 5-fold, XANES [23, 26];<br>5.1, XANES [27];<br>4.8-5.1, XRD [47] | 4.65, FF MD [49] |
| Mg-O | 4.18-4.84 | 4.30-4.89 | 5, XANES [70];<br>6, NMR [73, 75] | 4.5 RMC refinement [46]; 5.5, FF MD [66]; 4.74-5.09 RMC refinement [44, 52] |
| Ca-O | 5.99-6.67 | 6.08-6.74 | 6.9, NMR [73]; 7, XANES [78];<br>6-7, XANES [72]; 6.15, XRD [42] | 6.7, FF MD [66];<br>7-7.5, FF MD [78];<br>6.3, FF MD [49] |
| Na-O | 7.11-7.92 | 7.11-8.02 | 6, XANES [78]; | 6-6.5, FF MD [78]; |
| | | | 4.1-6.1, X-ray and neutron diffraction [62] | 5.1-5.3, *ab initio* MD [61];<br>4.4-5.1, FF MD [63];<br>4.8-5.3, FF MD [64] |
| K-O | 8.42-11.56 | 8.39-11.64 | 5.7-8.0, X-ray and neutron diffraction [62]<br>9-10, X-ray scattering [43] | 7.6-9.2, FF MD [65];<br>7.4-8.0, FF MD [63];<br>7.4-8.0, FF MD [64] |

## 3.2 Structural descriptors based on MD simulation results at room temperature

Based on the MD simulation results at room temperature presented in the previous section, we have calculated several structural descriptors in this section to evaluate their performance in describing dissolution rate data collected on the ten glasses at both acidic and alkaline conditions and far-from-equilibrium steady-state (see the data in Table 1, obtained from reference [7]).

### 3.2.1 Average metal-oxygen bond strength

We first introduce a structural descriptor, the average metal-oxygen (M-O) bond strength parameter, $S_{M-O}$, which is derived based on classical bond valence models, as illustrated in Equation (2).

$$S_{M-O} = \frac{\sum n_M \cdot s_M \cdot CN_M}{\sum n_M} \tag{2}$$

where $n_M$, $CN_M$ and $s_M$ are molar quantity, average oxygen coordination number (CN), and average bond valence of each type of metal cation M (M = Si, Al, Ti, $^{II}$Fe, $^{III}$Fe, Mg, Ca, Na and K). The bond valence $s_M$ is a measure of bond strength and commonly described as a function of bond length using Equation (3) [80].

$$s_M = s_{0_M} \left( \frac{R_{M-O}}{R_{0_{M-O}}} \right)^{-N_M} \tag{3}$$

where $R_{M-O}$ is the M-O bond length, and $s_{0_M}$, $R_{0_{M-O}}$, and $N_M$ are empirical parameters for each pair of atoms M-O, which have been determined by refining the model against accurate crystal structures as has been tabulated in reference [80]. The adopted values for these empirical parameters are given in Table S5 of the Supplementary Material.

Based on Equations (2) and (3) and the MD simulation results for the average CN (Table S4 of the Supplementary Material) and M-O bond distances (Table 4), we have calculated the average M-O bond strength parameter for each glass composition studied here. The results are presented in Figure 6 as a function of the log dissolution rate data (normalized by the BET surface area) collected on the same glasses at both alkaline (pH = 10.6) and acidic (pH = 4.0) conditions and far-from-equilibrium steady-state (data obtained from reference [7]). Figures 6a and 6b show the results based on glass composition in Tables 2 and 3, where all the Fe atoms are assumed to be 100% $^{II}Fe$ and $^{III}Fe$, respectively. The average M-O bond strength data in Figure 6c are then calculated using the values in Figure 6a-b by accounting for the measured ratio of $^{II}Fe/^{III}Fe$ (i.e., $FeO/Fe_2O_3$) for the ten glasses studied here (Table 1). It is clear from Figure 6c that the log dissolution rates are inversely correlated with the average M-O bond strength parameter, with $R^2$ values of 0.80-0.92 for linear regression. This inverse correlation is consistent with expectation because an oxide glass with a higher average M-O bond strength value means that, on average, it is harder to break/dissolve the glass, leading to a lower dissolution rate. Comparing the results in Figure 6a and 6b (see a direct comparison of the values in Fig. S4 of the Supplementary Material) reveals that the former (i.e., 100% $^{II}Fe$ composition) exhibits lower average M-O bond strength values than the latter (i.e., 100% $^{III}Fe$). This observation is attributed to the significantly lower bond valence value of $^{II}Fe$-O ($s_M$ = 0.42) compared with $^{III}Fe$-O ($s_M$ = 0.69). Nevertheless, we observe the same inverse correlation with similar $R^2$ values for linear regression in Figure 6a-c, which, together with the minimal standard deviations (see the error bars in Figure 6b) obtained from three independent MD production runs, are indications of the robustness of the method.

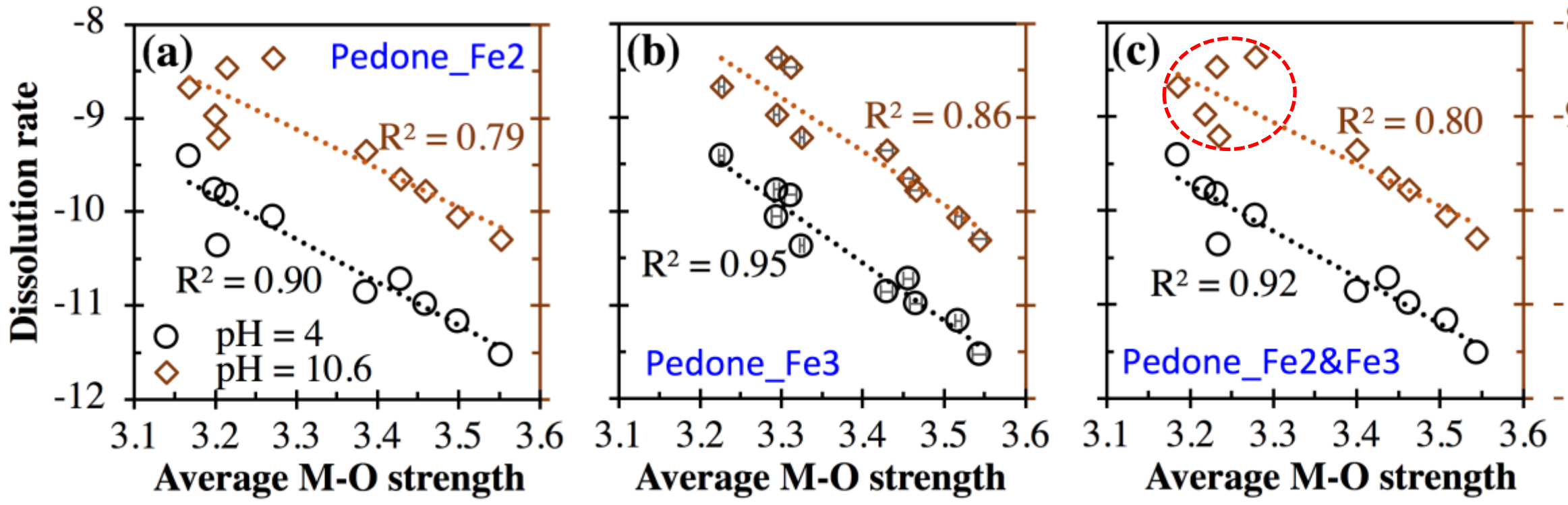

Figure 6. Correlation between (i) log Si dissolution rates (mol/m$^2$/s; normalized by BET surface area) of the ten volcanic glasses in Table 1 in solutions of pH 4.0 and 10.6 at far-from-equilibrium steady-state (data extracted from reference [7]) and (ii) the average metal-oxygen (M-O) bond strength, $S_{M-O}$, calculated using Equations (2) and (3). The $S_{M-O}$ values in (a) and (b) are determined from MD simulation results for glass compositions in Tables 2 and 3, where all the Fe atoms are assumed to be 100% $^{II}$Fe and $^{III}$Fe, respectively. The $S_{M-O}$ data in (c) are then estimated from the corresponding $S_{M-O}$ values in (a) and (b) by taking into account the measured molar ratio of $FeO/Fe_2O_3$ for each glass in Table 1. The acid solutions (pH = 4.0) were prepared using Merck analytical grade $NH_4Cl$ and HCl, while the alkaline solutions (pH = 10.6) were prepared using Merck analytical grade $NH_4Cl$ and $NH_3$ [7]. The error bars in (b) are one standard deviation based on the results of three independent MD production runs. The $R^2$ values for linear regression are shown in the figure.

Another important observation in Figure 6 is that the average M-O bond strength parameter appears to give a better description of glass dissolution rates in acidic (pH = 4) than alkaline (pH = 10.6) condition, as evidenced by the higher $R^2$ values for linear regression (0.90-0.95 versus 0.79-0.86), especially in the basaltic region with relatively low average M-O bond strength values (see the circled region in Figure 6c). The lower predictive performance for the glass dissolution rate data at pH = 10.6 in this region could be attributed to the fact that the calculation of the average M-O bond strength parameter (and any other form of structural descriptors or compositional parameters) assumes that all oxide components in the glass dissolve congruently (i.e., proportional to the glass composition). However, this assumption may deviate from the actual experiments in reference [7], where it has been shown that the dissolution of Si and Al atoms are incongruent. Although no data has been collected regarding the dissolution rate of other atoms (e.g., Ti, Fe, Ca, Mg, Na and K) in reference [7], an experimental investigation on crystalline basalt dissolution revealed significant deviations from congruent dissolution at pH of 10-11, especially for Fe and Mg atoms with respect to Si atom [81]. Specifically, we found from reference [81] that the $(Fe/Si)_{dissolved}/(Fe/Si)_{mineral}$ and $(Mg/Si)_{dissolved}/(Mg/Si)_{mineral}$ ratios at a temperature of 25 °C and pH of 10-11 are about ~0.004-0.18 and 0.01-0.18, respectively, which are significantly lower than one (i.e., congruent dissolution). In contrast, these ratios are close to one (hence congruent dissolution) for the dissolution data at a pH of 4 (i.e., 1.42-1.21). Therefore, it is possible that the glasses in the basaltic region (i.e., average M-O bond strength value of 3.2-3.3 in Figure 6), which have higher Fe and Mg content (~12-15 wt. % and ~3-6 wt. %) than the glasses in the rhyolitic region with higher average M-O bond strength values (~1-7 wt. % and ~0-2 wt.%), may have experienced a larger deviation from the assumption of congruent dissolution at the alkaline condition (pH = 10.6) than the acidic condition (pH =4). This difference may explain the lower predictive performance of the average M-O bond strength parameter in the basaltic region at the pH of 10.6.

Another widely used expression (a more recent development compared with Equation (2)) to describe the relationship between bond length $R_M$ and bond valence $s_M$ is given by Equation (4) [82].

$$s_{M-O} = e^{(R_{0_{M-O}} - R_M)/b} \quad (4)$$

where $b$ is commonly taken as a 'universal' constant equal to 0.37 [83], and $R_{0_M}$ is an empirical parameter that has been tabulated for different M-O pairs in reference [82]. The adopted $R_{0_M}$ values are given in Table S6 of the Supplementary Material.

The average M-O bond strength values calculated based on Equations (2) and (4) are compared with those based on Equations (2) and (3) in Figure 7. It is seen that the values obtained from Equations (2) and (4) are consistently lower than those obtained from Equations (2) and (3) by about 0.2. Nevertheless, these two parameters are seen to be linearly correlated with an $R^2$ value of ~1.00 for linear regression (Figure 7). As a result, the average M-O bond strength parameter based on Equations (2) and (4) exhibit a similar level of predictive performance for the glass dissolution rate data presented in Figure 6, as shown in Fig. S5 of the Supplementary Material.

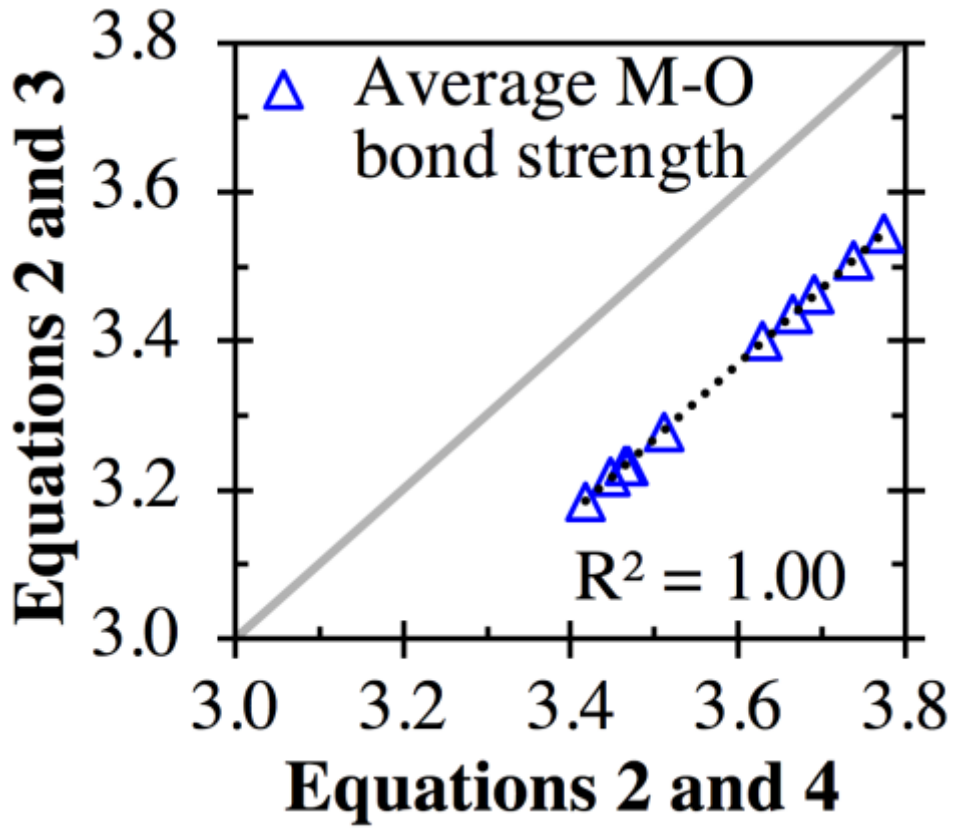

Figure 7. Correlation between the average M-O bond strength parameter calculated based on Equations (2) and (4) and those based on Equations (2) and (3). The $R^2$ value for linear regression is shown in the figure. The grey line in the figure is the line of equality.

### 3.2.2 Average metal oxygen dissociation energy

Next, we evaluate the performance of the so-called average metal oxide dissociation energy (AMODE) parameter in describing the same dissolution data, which has been introduced in a previous investigation to investigate the reactivity of C(M)AS glasses in alkaline environments [34]. The AMODE parameter is defined according to Equation (5).

$$AMODE = \frac{\sum N_M \cdot CN_M \cdot E_{M-O}}{\sum N_M} \tag{5}$$

where $N_M$ is the number of each type of metal cation ($M$ = Si, Al, Ti, Fe, Mg, Ca, Na, and K) in the oxide glass, $CN_M$ and $E_{M-O}$ are the average coordination number and the average energy required to break the single metal-oxygen bond for each type of metal cation $M$, respectively. The $CN_M$ values have been calculated from the MD simulations (Table S4 of the Supplementary Material), while the $E_{M-O}$ values can be obtained from the literature [84]. According to references [34, 84], the $E_{M-O}$ of the single Si-O, Mg-O, Ca-O, Ti-O, Na-O and K-O bonds are estimated as ~106, ~37, ~32, ~73, ~20 and ~13 kcal, respectively. For Al-O single bond, an $E_{M-O}$ value of 90, 75 and 60 kcal are adopted for IV-, V- and VI-fold Al atoms, respectively [34, 84]. For $^{II}$Fe and $^{III}$Fe, the corresponding $E_{M-O}$ value is estimated as 38 and 72 kcal, based on V- and IV-fold coordination states, respectively [85].

The AMODE values calculated using Equation (5) are presented in Figure 8 as a function of log dissolution rate at both alkaline and acidic conditions (similar to those presented in Figure 6). The results show inverse correlations between the AMODE parameter and the log dissolution rate data at both alkaline and acidic conditions. These inverse correlations are consistent with expectation since a lower AMODE value is a reflection of lower energy required to break/dissolve an oxide glass and hence is associated with a higher dissolution rate. Compared with the average M-O bond strength parameter derived based on classical bond valence models in Figure 6 and Fig. S5 of the Supplementary Material ($R^2$ = 0.79-0.95), the AMODE parameter in Figure 8 exhibits slightly lower $R^2$ values (0.76-0.91) for linear regressions, which suggests that the former slightly outperforms the AMODE parameter in capturing the log dissolution rates of these complex volcanic glasses. This slightly better performance of the former could be attributed to the accommodation of the potential deviation of single M-O bond strength in glasses from that in crystals by using the well-established bond valence models in Equation (3) and (4), where bond valence (or bond strength) is treated as a function of bond length. In contrast, the $E_{M-O}$values used to calculate the AMODE parameter in Equation (5) are estimated based on a single coordination state, which deviates from the CN results from MD simulations, especially for metal cations with relatively low field strength (e.g., Na and K, as seen in Figure 4).

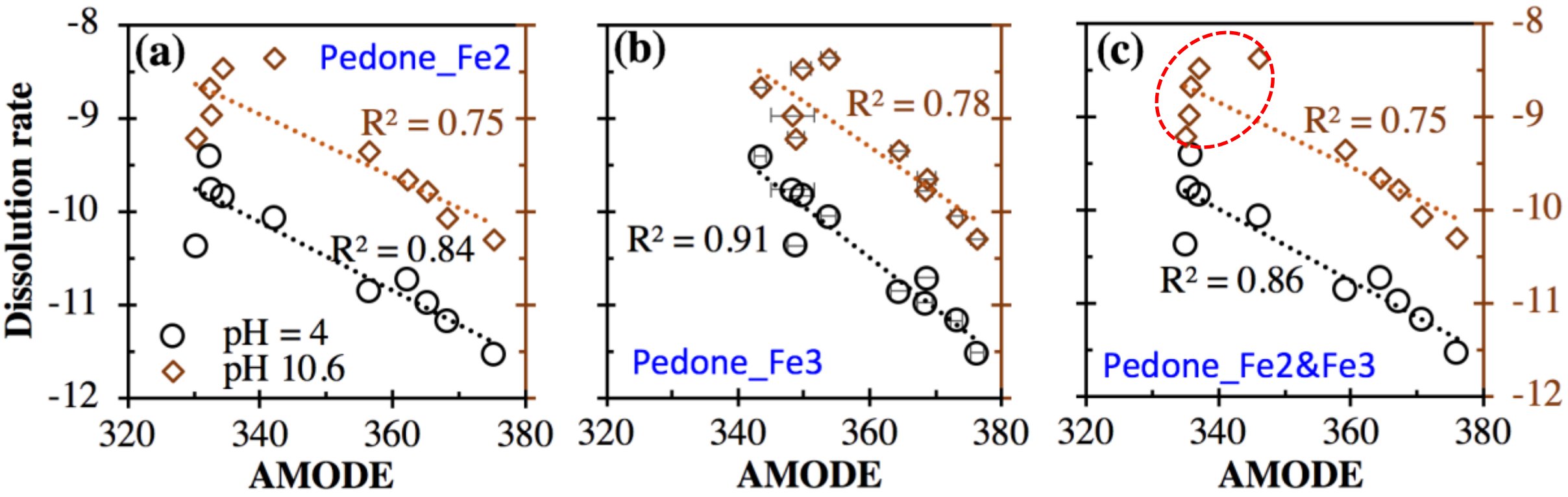

Figure 8. Correlation between (i) log Si dissolution rates (mol/m$^2$/s; normalized by BET surface area) of the ten volcanic glasses in Table 1 in solutions of pH 4.0 and 10.6 at far-from-equilibrium steady-state (data extracted from reference [7]) and (ii) the average metal oxide dissociation energy (AMODE) parameter calculated using Equation (5). The AMODE values in (a) and (b) are determined from MD simulation results for glass compositions in Tables 2 and 3, where all the Fe atoms are assumed to be 100% $^{II}$Fe and $^{III}$Fe, respectively. The AMODE values in (c) are then estimated from the values in (a) and (b) by taking into account the measured molar ratio of $FeO/Fe_2O_3$ in each glass in Table 1. The error bars in (b) are one standard deviation based on the results of three independent MD production runs. The $R^2$ values for linear regression are shown in the figure.

Furthermore, as expected, it is seen that the $^{III}$Fe glass composition (Figure 8b) give slightly larger AMODE values than the $^{II}$Fe glass composition (Figure 8a), similar to the case of the average M-O bond strength parameter (Figure 6a and 6b), which is attributed to the higher $E_{M-O}$ value of $^{III}$Fe (72 kcal) than $^{III}$Fe (38 kcal). A direct comparison of the AMODE values in Figure 8a and 8b is given in Fig. S6 of the Supplementary Material. The error bars in Figure 8b represent one standard deviation based on the analysis of results from three independent MD production runs. Relatively small standard deviations are indications of the robustness of the MD simulations and reproducibility of the AMODE parameter. Finally, we also observe that the AMODE parameter does not perform well in the basaltic region at the pH of 10.6, as highlighted by the dashed circle in Figure 8c. This observation is consistent with that of the average M-O bond strength parameter in Figure 6c. It could be attributed to the potentially larger deviation from congruent dissolution (congruent dissolution is the underlying assumption of all structural descriptors) for the basaltic composition at pH of 10-11 [81] compared to the (i) rhyolitic glass composition with relatively high AMODE values and (ii) glass dissolution data at the pH of 4.0, as has already been discussed in Section 3.2.1.

## 3.3 Structural descriptors based on dynamic properties

### 3.3.1 Average self-diffusion coefficient at melting

Another structural descriptor that has been shown previously to give predictive performance comparable to that of the AMODE parameter for C(M)AS glass dissolution is the so-called average self-diffusion coefficient (ASDC) at temperatures above the melting point [34]. This ASDC parameter gives an overall estimation of the mobility of atoms in the glasses, which to some extent reflects the easiness of breaking metal-oxygen bonds, in a sense similar to the bond-breaking process during glass dissolution. The ASDC parameter is calculated from mean square displacement (MSD) data (as shown in Figure 9) using Einstein's equation (i.e., Equation (6)).

$$D = \frac{\langle [r(t)-r(0)]^2 \rangle}{6t} \tag{6}$$

where D is the self-diffusion coefficient, $t$ is the simulation time, and $\langle[r(t) - r(0)]^2\rangle$ is the MSD between time $t$ and 0.

Figure 9 shows the MSD as a function of time for a typical glass (i.e., 13HEI_Fe2) at 2000 K calculated from the MD trajectories of the *NVT* equilibration step at 2000 K. By fitting the linear proportion of the MSD data (50-300 ps are selected here) using Equation (6), we can calculate the ASDC over all atoms in the glass (denoted as "all" in Figure 9) as well as for each type of atom. It is clear from Figure 9 that network formers (e.g., Si, Al and Ti) generally exhibit much lower mobility (smaller MSD at a given time) than network modifiers (e.g., $^{II}$Fe, Mg, Ca, Na and K), which is expected since Si-, Al- and Ti-O bonds are much stronger (hence harder to break) than $^{II}$Fe-, Mg-, Ca-, Na-O and K-O bonds.

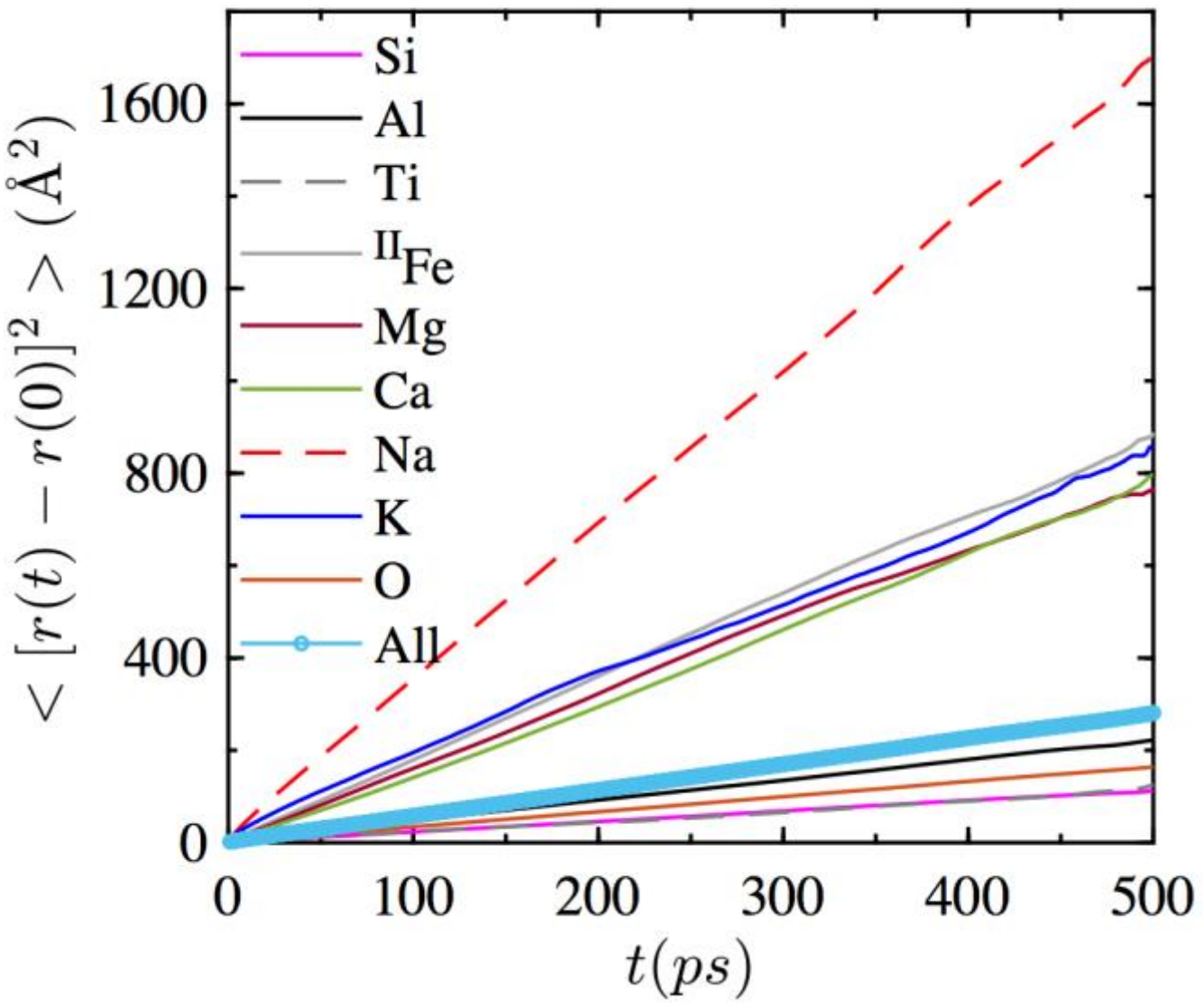

Figure 9. Mean square displacement (MSD) of each element along with the average of all atoms denoted as "All" in a typical glass (i.e., 13HEI_Fe2 in Table 2) as a function of time during the 500 ps of MD equilibration step at 2000 K.

Figure 10 shows how the ASDC parameter at 2000 K (after accounting for the measured molar ratio of $FeO/Fe_2O_3$ in each glass in Table 1) correlates with the same log dissolution rate data presented in the previous sections (i.e., Figures 6 and 8). The corresponding data for the 100% $^{II}$Fe and $^{III}$Fe glass composition in Tables 2 and 3 (similar to Figures 6a-b and 8a-b) are given in Fig. S7 of the Supplementary Material. All the results in Figure 10 and Fig. S7 show the expected positive correlations, with a higher ASDC value (hence higher atomic mobility) leading to generally higher dissolution rates. However, the $R^2$ values achieved with ASDC (0.60-0.76) are noticeably lower than those of (i) the average M-O bond strength parameter (0.80-0.92, Figure 6c) and (ii) the AMODE parameter (0.75-0.86, Figure 8c). This lower predictive performance of the ASDC parameter could be attributed to the fact that the mobility of metal cation is not only determined by the easiness to break the bonds formed with the nearby oxygen atoms but also the size of the metal cation and the availability of free space. This can be illustrated by comparing the MSD value of K and Na atoms in Figure 9, which shows that the slope of the MSD of the K atom (and hence the corresponding ASDC) is significantly lower than that of the Na atom, although the average Na-O bond strength (~0.11) is higher than that of K-O (~0.06) according to Equations (3)-(4). This contradiction can be mainly attributed to the larger ionic size of K (1.38 Å) than Na (1.00 Å) that renders the former more difficult to move around at 2000 K [86].

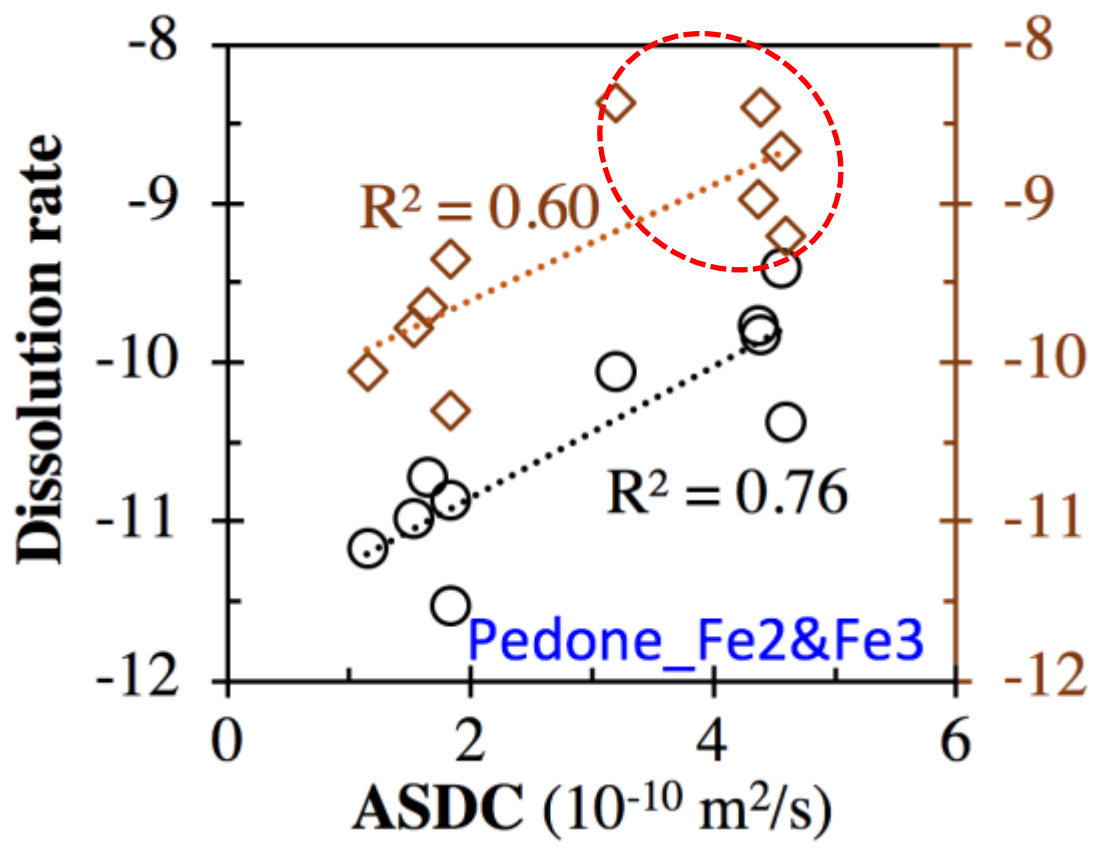

Figure 10. Correlation between (i) log Si dissolution rates (mol/m$^2$/s; normalized by BET surface area) of the glasses in Table 1 in solutions of pH 4.0 and 10.6 at far-from-equilibrium steady-state (data extracted from reference [7]) and (ii) the average self-diffusion coefficient

(ASDC) of all the atoms in the volcanic glasses at a temperature of 2000 K. The ASDC values in the figure have been calculated from the ASDC values in Fig. S7a-b of the Supplementary Material by taking into account the measured molar ratio of $FeO/Fe_2O_3$ in each glass in Table 1. More details on the dissolution data have been given in the figure caption of Figure 6.

The inability to adequately capture the data in the basaltic region with relatively higher ASDC is also observed for the ASDC parameter in Figure 10 (highlighted by the dashed circle), which is consistent with the two structural descriptors based on room temperature MD results (Section 3.2). As discussed in Section 3.2.1, this could be associated with larger deviations of basaltic glasses from congruent dissolution (especially in alkaline conditions) [67], which is the underlying assumption for all structural descriptors.

### 3.3.2 Energy barrier of self-diffusion

By calculating the self-diffusion coefficients (using Equation (6)) based on the MD trajectories of the *NVT* equilibration step at different temperatures (i.e., 2000, 3000, and 5000 K), it is possible to estimate the energy barrier for self-diffusion by using the Arrhenius relationship of diffusion coefficient as shown in Equation (7) [87].

$$\ln(D) = \ln(D_0) - \frac{\Delta E_a}{RT} \tag{7}$$

where $\Delta E_a$ is the self-diffusion energy barrier, $T$ is the temperature, and $R$ is the gas constant.

Figure 11 shows the correlation between $\ln(D)$ and 1000/T for a typical glass composition (i.e., 13HEI in Tables 2 and 3), from which the slope of the curve is directly proportional to $\Delta E_a$ in Equation (7), via $\Delta E_a = -1000R \times slope$. From the slopes in Figure 11, we see that the energy barrier of diffusion for the $^{III}$Fe glass (116 kJ/mol) is higher than that of the $^{II}$Fe glass (110 kJ/mol). This observation is consistent with the results in the previous sections and can be attributed to the higher overall average $^{III}$Fe-O bond strength (~0.76) and $^{III}$Fe-O bond dissociation energy (~72 kcal) than those of $^{II}$Fe-O (~0.45 and ~38 kcal, respectively). These MD-derived energy barriers are seen to be higher than the apparent activation energy of the same glasses calculated using dissolution rate data at different temperatures (i.e., 27-56 kJ/mol, obtained from reference [7]). The energy barrier for the self-diffusion of Na atom in this glass (i.e., 13HEI) is estimated to be ~70-77 kJ/mol depending on the type of iron ($^{II}$Fe versus $^{III}$Fe), which is in reasonable agreement with previous MD simulations (72 kJ/mol [87] and 85 kJ/mol [88]) and experimental data (75.5 ± 1.2 kJ/mol [89]) on sodium aluminosilicate glasses.

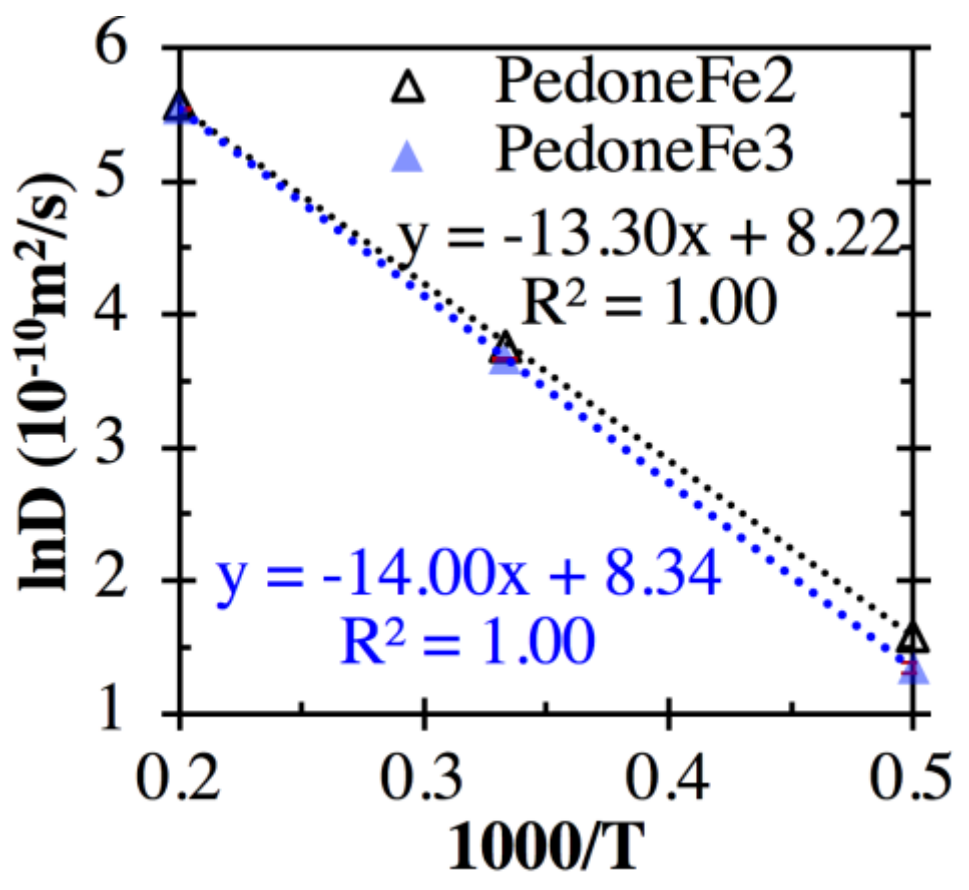

Figure 11. Correlation between $\ln(D)$ and 1000/T for a typical glass (i.e., 13HEI). The self-diffusion coefficient $D$ at each temperature is obtained using Equation (6) based on the MSD of all the atoms in the glass. The error bars (in red) for the PedoneFe3 data are one standard deviation based on three independent MD production runs. Also given in the figure are $R^2$ values and equations of fit for linear regressions.

Figure 12 shows how the energy barrier of self-diffusion (after accounting for the measured molar ratio of $FeO/Fe_2O_3$ in each glass in Table 1) correlates with the same log dissolution rate data presented in previous sections. The results for 100% $^{II}$Fe and $^{III}$Fe glass compositions in Tables 2 and 3 (similar to Figures 6a-b and 8a-b) are given in Fig. S8 of the Supplementary Material. All the results in Figure 12 and Fig. S8 show the expected inverse correlations with a higher energy barrier leading to generally lower dissolution rates. Compared with the two structural descriptors presented in Section 3.2, the energy barrier of self-diffusion gives much lower $R^2$ values for linear regression, similar to the case of ASDC at 2000 K. The main reason for the relatively low predictive performance of these descriptors based on dynamic properties (as has been discussed in Section 3.3.1) is that it is influenced not only by the easiness to break the bonds formed with the nearby oxygen atoms but also the size of the metal cation. For example, K cation is seen to have a much larger energy barrier of self-diffusion (~89-94 kJ/mol) than Na cation (~70-77 kJ/mol) although the former has lower M-O bond strength and bond dissociation energy.

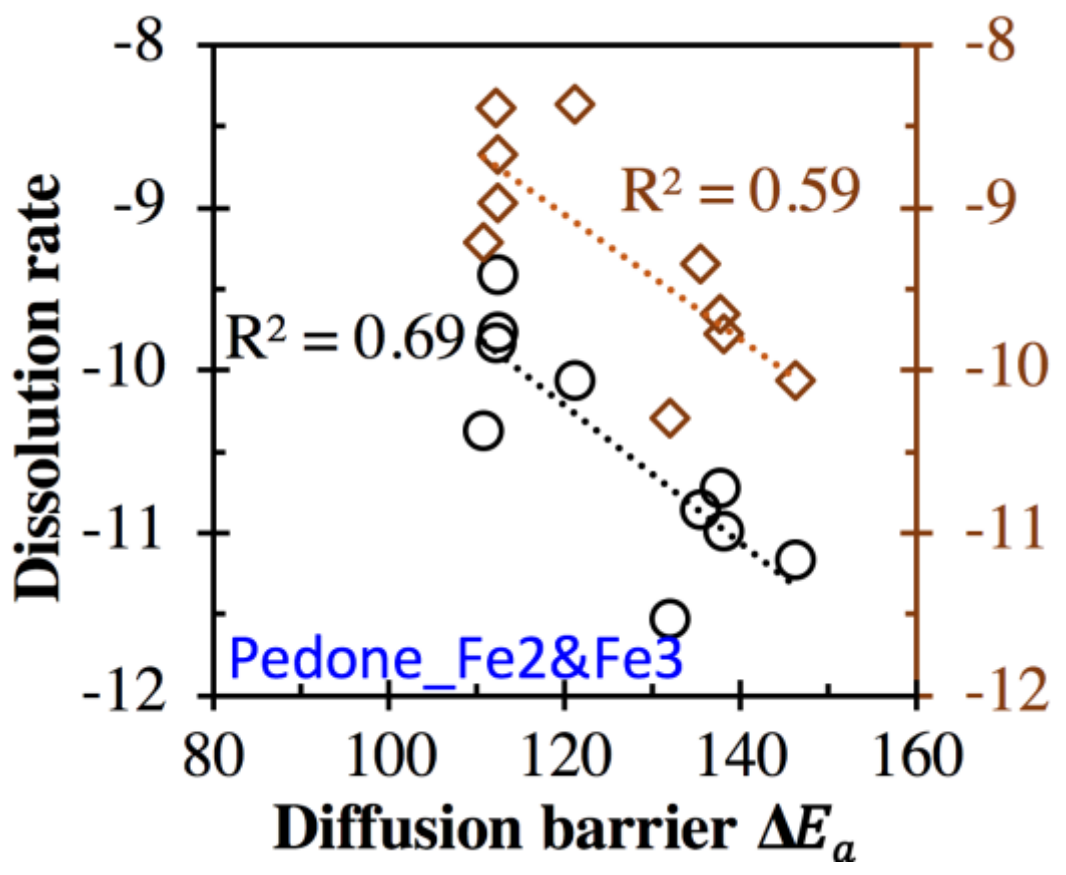

Figure 12. Correlation between (i) log Si dissolution rates ($mol/m^2/s$; normalized by BET surface area) of the glasses in Table 1 in solutions of pH 4.0 and 10.6 at far-from-equilibrium steady-state (data extracted from reference [7]) and (ii) the energy barrier of self-diffusion, $\Delta E_a$, calculated from the MD simulation results using Equations (7). The $\Delta E_a$ values in the figure have been calculated from the $\Delta E_a$ values in Fig. S8a-b of the Supplementary Material by taking into account the measured molar ratio of $FeO/Fe_2O_3$ in each glass in Table 1.

### 3.4 Comparison with the extent of depolymerization parameter

Finally, the commonly used extent of depolymerization (i.e., NBO/T) has been calculated based on the glass composition in Table 1 following reference [20] (details given in Section 14 of the Supplementary Material). Figure 13 shows the correlation between the NBO/T parameter and log dissolution rates of the ten glasses studied here (similar to those presented in Figures 6, 8, 10 and 12). The expected overall positive correlation is observed, with a higher NBO/T value leading to a higher dissolution rate. However, the level of correlation achieved with the NBO/T parameter is generally lower than the descriptors presented in Sections 3.2-3.3, especially for the average M-O bond strength and AMODE parameters, as illustrated by the comparison of the $R^2$ values for linear regressions in Table 6. The generally better predictive performance of the MD-derived structural descriptors is attributed to the fact that they better distinguish the difference among different modifiers and network formers than NBO/T (only distinguish network formers from modifiers), especially for the average M-O strength and AMODE parameters, where the easiness of breaking each type of M-O bond is differentiated. The average M-O bond strength parameter derived in Section 3.2.1 based on classical bond valence models exhibit the best overall predictive performance (slightly higher $R^2$ values when compared with the AMODE parameter, as shown in Table 6), which has been attributed to its consideration of single M-O bond strength as a function of M-O bond length. These observations demonstrate that the average M-O bond strength parameter introduced here is also a promising structural descriptor to predict the dissolution rates of highly complex multicomponent silicate glasses from their chemical composition. Furthermore, the empirical parameters in the bond valence models (e.g., Equations (3) and (4)) have been well tabulated for a wide range of chemicals (not just oxides) [80, 82], suggesting possible broader applications for the average M-O bond strength parameters (in comparison with the AMODE parameter).

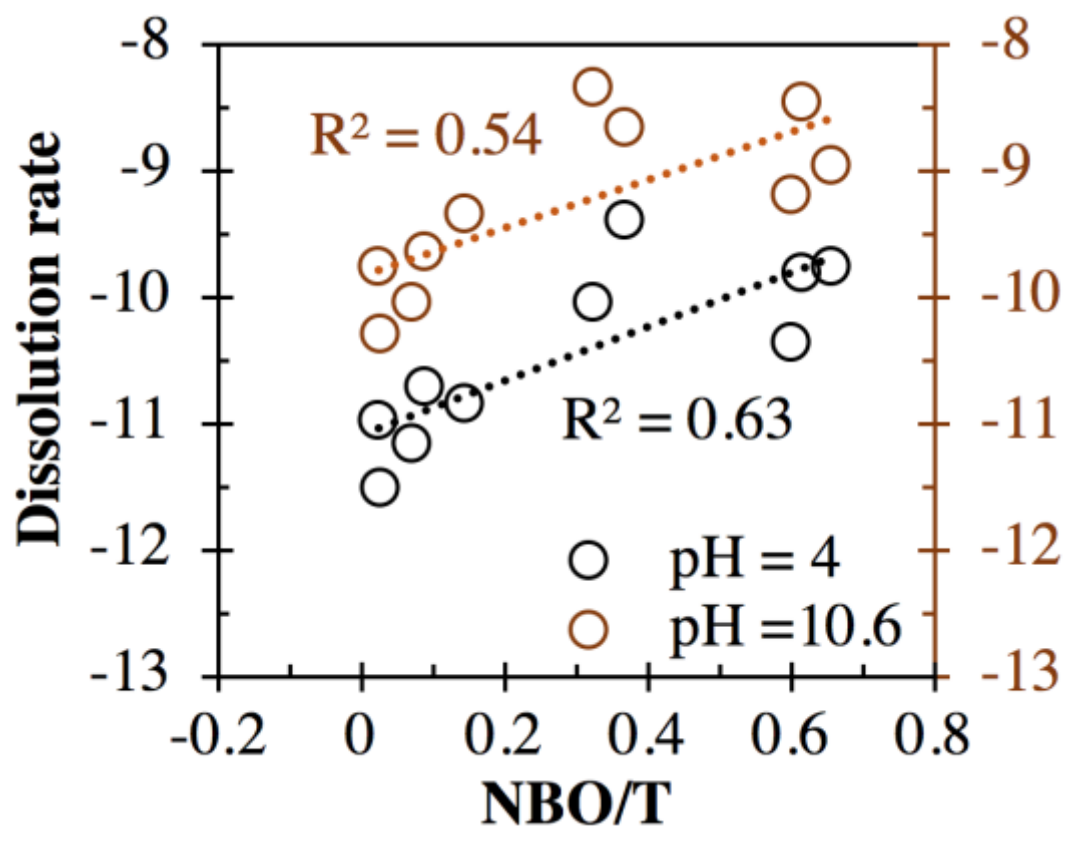

Figure 13. Correlation between the extent of depolymerization (NBO/T) and log Si dissolution rate of the ten complex natural glasses at (a) pH = 4 and (b) pH =10.6 solutions (dissolution data obtained from reference [7]).

Table 6. Summary of the level of correlation ($R^2$ values) achieved for linear regressions between the different structural descriptors and the log Si dissolution rates of the ten natural volcanic glasses in both basic and acidic pH (see Figures 6, 8, 10, 12 and 13).

| Structural descriptors | $R^2$ values achieved for linear regression | |
|---|---|---|
| | pH = 4 | pH =10.6 |
| Average M-O bond strength: Equation (2) | 0.92 | 0.80 |
| Average M-O bond strength: Equation (3) | 0.92 | 0.80 |
| AMODE | 0.86 | 0.75 |
| ASDC at 2000 K | 0.75 | 0.60 |
| Energy barrier of self-diffusion | 0.69 | 0.59 |
| NBO/T | 0.63 | 0.54 |

## 3.5 Limitations and broader impact

### 3.5.1 Limitations

As mentioned in the previous sections, a shared limitation of all the structural descriptors and compositional parameters that have been developed to describe glass dissolution/reactivity is that all these parameters assume congruent dissolutions for all the oxide components in the glasses. However, this assumption may not hold in the actual experimental conditions and the extent of deviation from congruent dissolution can vary significantly depending on the pH and temperature [81] and glass compositions [90]. This discrepancy between the actual experiments and the underlying assumption of congruent dissolution for these structural descriptors may have contributed to the poor predictive performance of all the parameters in the basaltic region at the pH of 10.6, as has been discussed in Section 3.2.

The dissolution of natural silicate glasses is highly complex, and several other factors could have a large impact on the dissolution rates, including particle size distribution, potential phase segregation, degree of amorphousness (and presence of crystalline phases), solution chemistry, pH and temperature [81], in addition to the glass composition and structure presented in this investigation. Furthermore, although MD simulations here are seen to generate reasonable structural representations for these highly complex natural glasses, there are still some discrepancies between the simulation results and experimental data, e.g., the CNs of Ti, Na and K (Table 5). MD simulations also bear several common limitations that have been briefly discussed in previous investigations [34, 39], including fast cooling rate and limited cell size (as compared to actual experiments/samples). Finally, each structural descriptor has its limitations, as has been briefly discussed in the previous section. Taking the best performing parameter (i.e., the average M-O bond strength parameter in Section 3.2.1) for example, the M-O bond length is obtained from the peak position of the partial RDFs, which represent the most probable bond distances, whereas the average coordination numbers are used in Equation (2). Given each metal cation M adopts a distribution of CNs and bond length, a possibly more accurate approach is to calculate the average M-O bond distances as a function of CN for each type of metal cation. This, however, would significantly increase the complexity of calculating this parameter and decrease its applicability, and hence is not explored in this article, but should be worth exploring in the future for simpler glasses.

#### 3.5.2 Broader impact

This investigation has demonstrated the ability of MD simulations (based on the Pedone force field [38] and the "melt-and-quench" method) to generate detailed structural information for highly complex CMASTFNK glasses at room temperature, which is still rare in the existing literature. Most existing MD studies on similar CMASTFNK glass systems focused on high temperature or pressure properties [35, 38], and it is experimentally challenging/expensive to obtain detailed coordination states for Ca, Mg, Na and K in such highly complex natural glasses. Second, we have extended the AMODE parameter previously derived for simpler quaternary CMAS and ternary CAS glasses to cover more compositionally complex glass systems and demonstrated the ability of the AMODE parameter of capturing the far-from-equilibrium dissolution rates for these natural volcanic glasses at both acidic and alkaline conditions. Furthermore, the newly introduced structural descriptors derived based on the classical bond valence models (i.e., average M-O bond strength parameter) have been shown to exhibit slightly better predictive performance (slightly higher $R^2$ values) than the AMODE parameter. Furthermore, in addition to the field geochemistry, dissolution (or chemical durability and reactivity) of glasses and minerals are important to study for many other fields and engineering applications, including nuclear waste encapsulation, bioglass dissolution, and carbon mineralization, blended cements and concrete, and alkaline activated materials. Hence, the structural descriptors and method of deviations can be potentially extended to those fields and applications to cover more glasses and minerals.

# 4 Conclusions

We employed force field-based molecular dynamic (MD) simulations with a commonly used force field covering the CaO-MgO-$Al_2O_3$-$SiO_2$-$TiO_2$-FeO-$Fe_2O_3$-$Na_2O$-$K_2O$ systems (developed by Pedone *et al.* [38]) to generate detailed structural representations for ten natural volcanic glasses of different compositions. Based on analysis of the resulting structural representations at 300 K, we have calculated several important structural attributes, including the radial distribution functions (RDFs), the nearest interatomic distances and coordination number (CN). Comparison of these structural attributes with those available in the literature on silicate-based glasses reveals that the nearest interatomic distances and the CNs of all the metal cations are captured reasonably well. Furthermore, the glass composition is seen to have a negligible impact on the nearest interatomic distances, but its impact on CNs is significant, especially for metal cations with low field strength (e.g., Na and K).

Based on analysis of the MD-generated structures at 300 K and classical bond valence models, we have introduced a novel structural descriptor, i.e., average metal-oxygen (M-O) bond strength parameter, which is shown to exhibit a strong inverse correlation with log dissolution rate data collected on these glasses at pH of 4 and 10.6 and far-from-equilibrium steady-state ($R^2$ values of ~0.80-0.92). The performance of this descriptor is shown to outperform several other descriptors also derived from MD simulation results, including average metal oxide dissociation energy (AMODE), average self-diffusion coefficient (ASDC) of all the atoms at melting, and energy barrier of self-diffusion. All the MD-derived descriptors are shown to give better predictive performance than the commonly used extent of depolymerization (NBO/T) parameter. The superior performance of the average M-O bond strength and AMODE parameters (especially the former) has been attributed to their better distinction of the easiness of breaking different types of M-O bonds. The results suggest that both parameters are promising structural descriptors to predict the dissolution rate (or reactivity) of highly complex natural glasses from their chemical composition. Hence, this study has highlighted the power of force field MD simulations in helping establish the important composition-structure-reactivity relationships for amorphous materials.

# 5 Acknowledgment

This material is based on work supported by MIT-IBM Watson AI Lab. The MD simulations were performed on computational resources supported by the Microsystems Technology Laboratories at Massachusetts Institute of Technology. The authors also acknowledge the discussions with Hugo Uvegi, Brian Traynor and Tunahan Aytaş.

# 6 Supplementary Material

Correlations between the quantities of oxide components; Force field parameters; Estimation of the glass density at different temperatures; Calculation of radial distribution functions; Nearest interatomic distances from MD simulations; Determination of cutoff distances; Summary of average coordination number; Average metal-oxygen bond strength parameter; AMODE parameter; Average self-diffusion coefficient at melting; Energy barrier of self-diffusion; Calculation of extent of depolymerization